\documentclass[11pt, a4paper]{article}
\usepackage{jcappub}
\usepackage{epsfig}
\usepackage{natbib}
\usepackage{amsmath}
\usepackage{amsfonts}
\usepackage{amssymb}
\usepackage{graphicx}
\usepackage{hyperref}
\usepackage{caption}
\usepackage{subcaption}
\usepackage{comment}
\usepackage{paralist}
\usepackage[shortlabels]{enumitem}
\usepackage[normalem]{ulem}
\usepackage[dvipsnames]{xcolor}

\title{ULDM self-interactions, tidal effects and tunnelling out of satellite galaxies}
\author[a]{Bihag Dave,}
\author[b, c]{and Gaurav Goswami}

\affiliation[a]{School of Engineering and Applied Science, Ahmedabad University, Commerce Six Roads, Navrangpura, Ahmedabad - 380009, India}
\affiliation[b]{Division of Mathematical and Physical Sciences, School of Arts and Sciences, Ahmedabad University, Commerce Six Roads, Navrangpura, Ahmedabad - 380009, India}
\affiliation[c]{International Centre for Space and Cosmology, Ahmedabad University, Commerce Six Roads, Navrangpura, Ahmedabad - 380009, India}

\emailAdd{bihag.d@ahduni.edu.in}
\emailAdd{gaurav.goswami@ahduni.edu.in}

\abstract{
It is well-known that Dark Matter (DM) inside a satellite galaxy orbiting a host halo experiences a tidal potential. If DM is ultra-light, given its wave-like nature, one expects it to tunnel out of the satellite - if this happens sufficiently quickly, then the satellite will not survive over cosmological timescales, severely constraining this dark matter model.  
In this paper, we study the effects of the inevitable quartic self-interaction of scalar Ultra-Light Dark Matter (ULDM) on the lifetimes of satellite galaxies by looking for quasi-stationary solutions with outgoing wave boundary conditions. 
For a satellite with some known core mass and orbital period, we find that, attractive (repulsive) self-interactions decrease (increase) the rate of tunnelling of DM out of it.  
In particular, for satellite galaxies with core mass $\sim \mathcal{O}(10^7-10^8)\ M_\odot$ and orbital period $\sim \mathcal{O}(1)\ \text{Gyr}$, one can impose constraints on the strength of self-interactions as small as $\lambda\sim \mathcal{O}(10^{-92})$. For instance, for ULDM mass $m = 10^{-22}\ \text{eV}$, the existence of the Fornax dwarf galaxy necessitates attractive self-interactions with $\lambda \lesssim -2.12 \times 10^{-91}$.
}


\arxivnumber{2310.19664}

\begin{document}
\maketitle
\flushbottom

\section{Introduction}\label{sec:Introduction}
Astrophysical observations of mergers of galaxy clusters \cite{Clowe:2003tk}, velocity dispersion of galaxies inside clusters \cite{Zwicky:1933gu}, and galactic rotation curves \cite{Rubin:1980orc} suggest that a significant fraction of mass within galaxies and clusters comes from a non-luminous, non-baryonic, cold, collisionless clustering component. Signatures of this component, called Dark Matter (DM), can also be seen at cosmological scales. For example, primordial light elemental abundances suggest that $\Omega_{b} $ is much smaller than $\Omega_{m}$ inferred from other observations while anisotropies in the Cosmic Microwave Background (CMB) imply a cold dark matter density fraction $\Omega_{c}h^2 \approx 0.12$ while the baryon density fraction is $\Omega_{b}h^2 \approx 0.02$ \cite{Planck:2018vyg, Profumo:2019ujg, Safdi:2022xkm}. 

If DM consists of an elementary particle, its mass, spin, interaction strengths etc. are largely unknown. From various experiments and observations, we know that its non-gravitational interactions with standard model particles are quite small \cite{ParticleDataGroup:2022pth, NANOGrav:2023hvm}. In this work, we examine the possibility that DM consists of spinless particles which are ultralight ($m \sim 10^{-22}\ \text{eV}$), whose inevitable quartic self-coupling ($\lambda$) is non-negligible. This kind of DM can exhibit wave-like behaviour, hence is described using classical field theory (see for instance \cite{Herdeiro:2022gzp, Eberhardt:2023axk} for some recent papers justifying this) and is referred to as Ultra Light Dark Matter (ULDM) and if $\lambda = 0$, Fuzzy Dark Matter (FDM) \cite{Ferreira:2020fam, Hui:2021tkt, 2168507}. In such scenarios, even very small values of self-coupling ($\lambda \sim 10^{-96}$ for $m\sim 10^{-22}\ \text{eV}$ \cite{Desjacques:2017fmf, Chakrabarti:2022owq}) can significantly impact the size and masses of stable configurations \cite{Colpi:1986psi, Chavanis:2011mrr}. Note that self-interactions are extremely important in another model of dark matter as well viz., Self-Interacting Dark Matter (SIDM) (see \cite{Tulin:2017ara} for a recent review). It is therefore an interesting question to ask how these self-interactions affect observational signatures of ULDM. 

\begin{figure}[ht]
    \centering
    \begin{subfigure}[b]{0.45\textwidth}
        \centering
        \includegraphics[width = \textwidth]{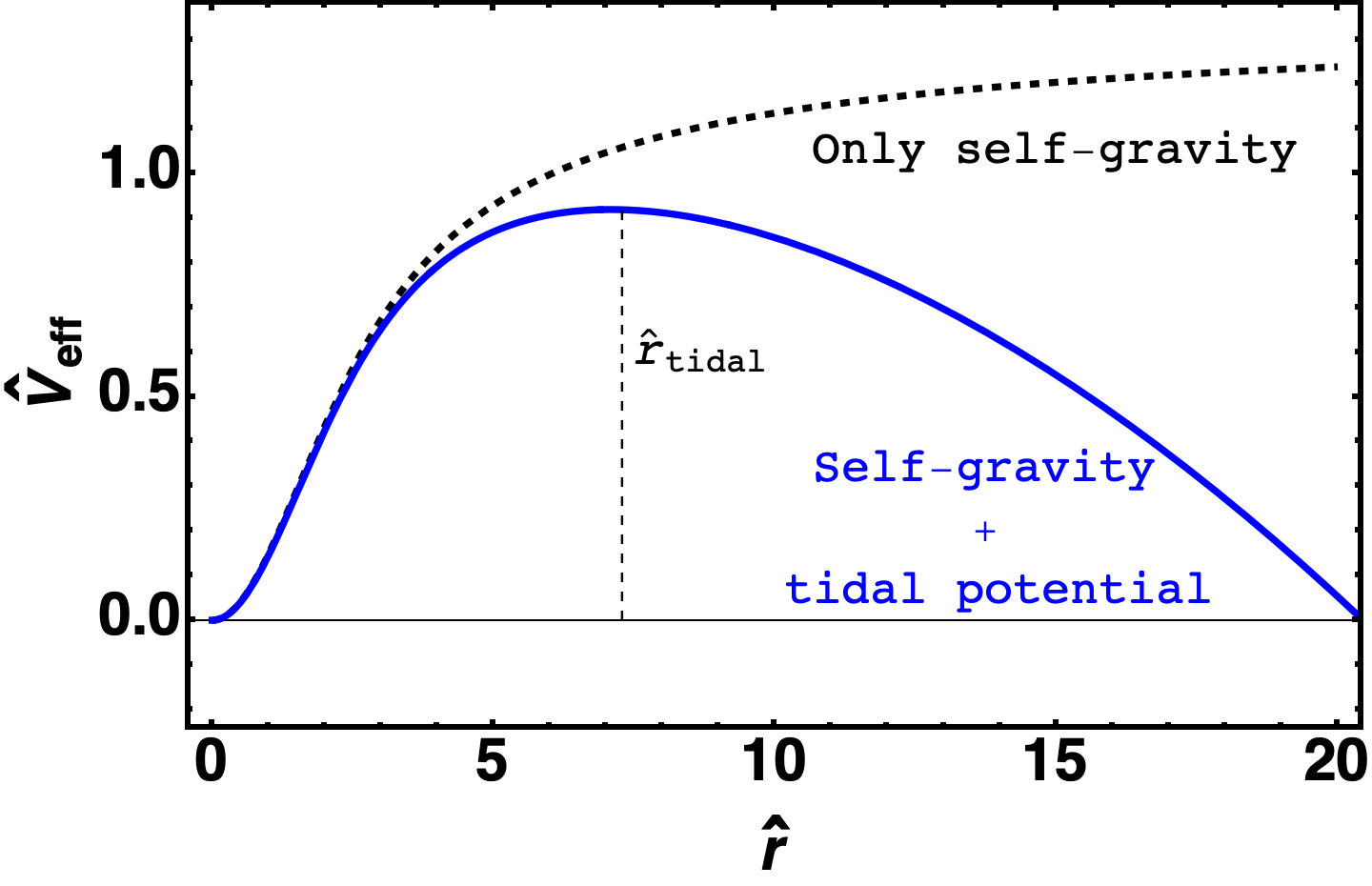}
        \caption{In the absence of a tidal potential (black dashed curve), there is no local maximum in effective potential while in the presence of an external tidal potential (blue), the effective potential has a peak at the tidal radius.}
        \label{fig:barrier}
    \end{subfigure}
    \hfill
    \begin{subfigure}[b]{0.45\textwidth}
        \centering
        \includegraphics[width = \textwidth]{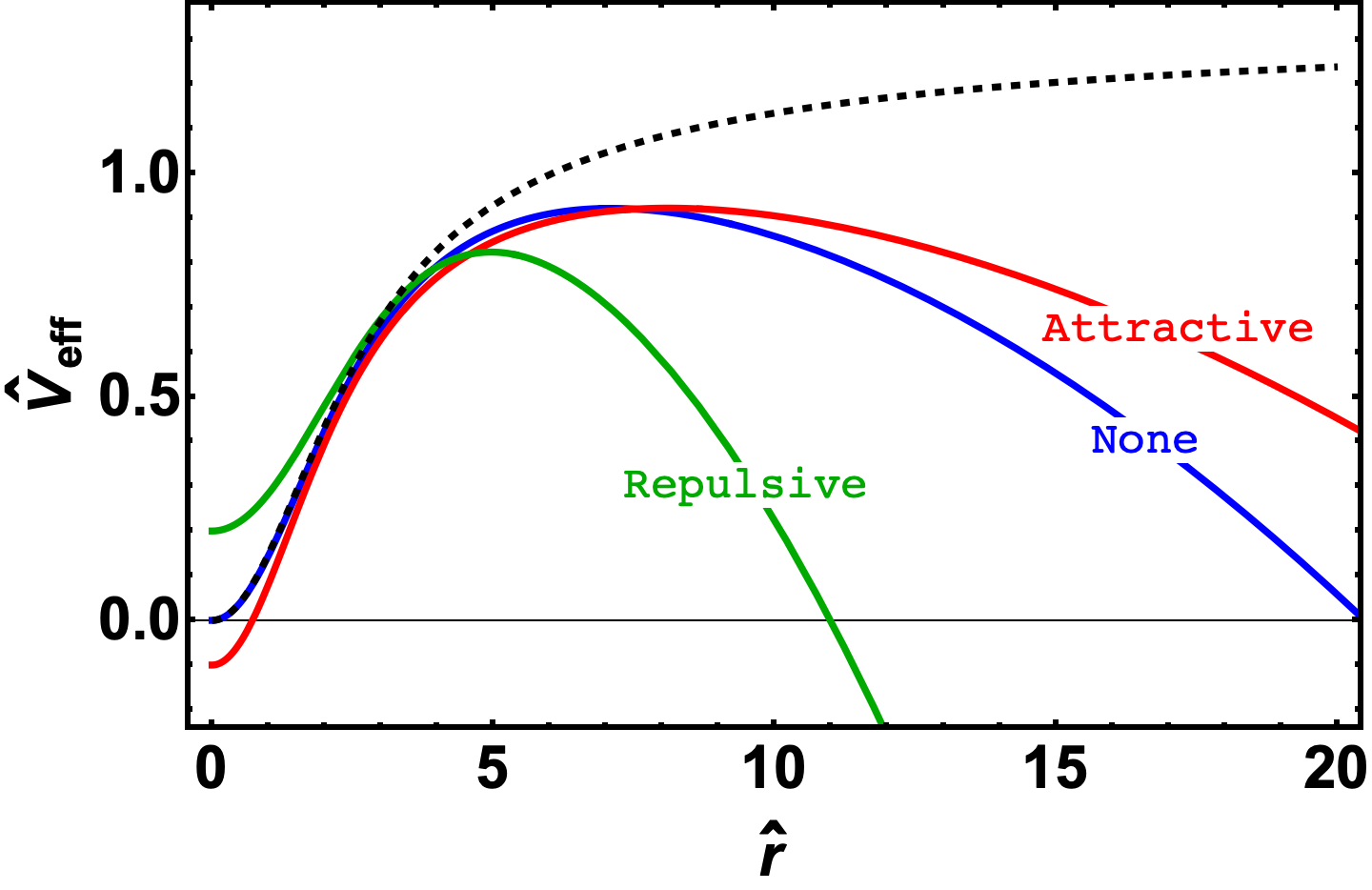}
        \caption{For an object with a fixed mass and orbital period (used in section~\ref{sec:fixed_Mc}), the potential barrier shrinks in the presence of repulsive self-interactions (green), while it stretches when self-interactions are attractive (red).}
        \label{fig:SI_barrier}
    \end{subfigure}
    \caption{Effective potential (eq.~(\ref{eq:effective_potential}), see also section~\ref{sec:numerical_procedure} for definitions of $\hat{r},~ \hat{V}_\text{eff}$) obtained after solving the modified Gross-Pitaevskii-Poisson equations (eqs.~(\ref{eq:mod_GP}) and~(\ref{eq:mod_poisson})) numerically. Left panel shows the effective potential for $\lambda = 0$, while the right panel shows the same for $\lambda \approx -2.9\times 10^{-92}$ (red), $\lambda \approx 1.4\times 10^{-91}$ (green). 
    }
    \label{fig:changing_barriers}
\end{figure}

We explore an interesting consequence of the wave-like nature of ULDM: the loss of mass from satellite dwarf galaxies due to tidal effects. Consider a satellite galaxy in a circular orbit around the centre of a larger host DM halo. In this system, self-gravity of the satellite holds it together, while tidal forces due to the host halo contribute a tidal potential that leads to a local maximum in the effective potential corresponding to a tidal radius $r_\text{tidal}$ (measured from the centre of the satellite galaxy) \cite{vandenBosch:2017ynq}. In this case, the effective potential resembles a barrier as shown in figure~\ref{fig:barrier} for $\lambda = 0$.

Note that in the case of particle-like Cold Dark Matter (CDM), matter contained within the tidal radius ($r < r_\text{tidal}$ in the rest frame of the satellite) will be safe from tidal disruption indefinitely. On the other hand, wave-like dark matter inside the tidal radius will tunnel out through the potential barrier over time, leading to the eventual tidal disruption of the satellite \cite{Hui:2016ltb} which can potentially address the `Missing Satellites' problem of CDM. If one finds that the rate of tunnelling for an observed satellite is too high to survive on cosmological timescales, one can impose constraints on the parameters of ULDM \cite{Du:2018qor, Hertzberg:2022vhk}. Note that these constraints are obtained by assuming negligible self-interactions, i.e. $\lambda = 0$. 

In this paper, we extend the problem to include self-interactions and ask: Do self-interactions of ULDM have a significant impact on the lifetimes of satellite galaxies? We find that the answer is yes, they do. As shown in figure~\ref{fig:SI_barrier}, we find that for an object with some known core mass and orbital period, attractive (repulsive) self-interactions lead to a wider (narrower) potential barrier which in-turn leads to a longer (shorter)-lived satellite. The results of our analysis also agree well with the simulations carried out recently by \cite{Glennon:2022huu} in the presence of self-interactions. 

Another way to see this is the following: We know that for a fixed soliton mass, central densities are higher for attractive self-interactions while lower for repulsive self-interactions (see figure~1 of \cite{Chakrabarti:2022owq}). We also know from \cite{Hui:2016ltb, Du:2018qor, Hertzberg:2022vhk} that the lifetime of a satellite with a fixed orbital frequency is longer (shorter) for a larger (smaller) central density (see section~\ref{sec:comparison} for details). In section~\ref{sec:fixed_Mc}, we confirm the above intuition. In particular, for a given ULDM mass $m$ and no self-interactions, if a satellite galaxy does not survive over cosmological timescales, then sufficiently strong attractive self-interactions allow it to survive given the same ULDM mass $m$.

In this paper, we model the loss of dark matter from a satellite galaxy by treating it as a tunnelling problem. We do so in two ways: 
(a) by looking for quasi-stationary solutions to a time-independent version of the equations of motion in the presence of a tidal potential by allowing the ``energy" to be complex (dealt with in the main body of the paper), and, (b) by using the usual textbook approach of dealing with tunneling phenomena with real ``energy" (discussed in the appendix). 
In the first approach, the solutions we deal with are quasi-stationary because in order to model a `leaky' satellite one must employ outgoing wave boundary conditions, leading to non-normalisable solutions whose density decreases with time (see section~\ref{sec:quasi-stationary} for more details). The rate of tunnelling $\Gamma$ is characterised by the imaginary part of the now complex energy eigenvalue while the lifetime of the satellite is defined by its inverse ($\tau \equiv 1/\Gamma$). To illustrate the effects of self-interactions on realistic satellite galaxies, we look at different satellites with the same known observable quantities. 
It is important to note that the effects of self-interactions appear differently depending on whether we look at satellites with the same central density and orbital period or with the same core mass and orbital period - this issue is discussed in great detail in sections~\ref{sec:fixed_rhoC} and~\ref{sec:fixed_Mc}.

The rest of the paper is organized as follows: In section~\ref{sec:modified_GPP}, we look at how the Gross-Pitaevskii-Poisson system is altered in the presence of an external tidal potential. Here we also discuss the quasi-stationary states, appropriate boundary conditions, and the rate at which a satellite galaxy would lose mass. In section~\ref{sec:numerical_procedure}, we solve the system numerically, obtain tunnelling rates and lifetimes as well as introduce scaling relations. In section~\ref{sec:SI_impact} we study the impact of self-interactions by looking at objects with fixed known properties (like central density or core mass). Finally in section~\ref{sec:summary} we summarize our results and discuss implications on future work. As mentioned above, in appendix~\ref{app:approx_appraoch}, we develop intuition for the change in potential barriers by repeating the analysis using a textbook approach similar to e.g. the $\alpha$-decay problem.

\section{Satellite galaxy in the presence of a tidal potential}\label{sec:modified_GPP}

Before we introduce a tidal potential for ULDM, we must first talk about the usual system of equations that we deal with in the absence of tidal effects. 
We begin with a classical real scalar field $\varphi$, which has the potential $U(\varphi) = \frac{1}{2}\frac{m^2c^2}{\hbar^2}\varphi^2 + \frac{1}{4!}\frac{\lambda}{\hbar}\varphi^4$, where $\lambda$ dictates the strength of self-interactions. 
Here, $\lambda < 0$ implies attractive self-interactions while $\lambda > 0$ implies repulsive self-interactions. 

In the non-relativistic limit, the real $\varphi$ can be written in terms of a complex field, $\varphi = \frac{\hbar}{m\sqrt{2c}}\left(e^{-imc^2t/\hbar}\Psi + c.c.\right)$, and by averaging out the rapidly oscillating modes and taking the weak-gravity limit, the scalar field can be described using the Gross-Pitaevskii-Poisson equations (see \cite{Chakrabarti:2022owq} for a detailed derivation),

\begin{eqnarray}
    i\hbar\frac{\partial}{\partial t}\Psi &=& -\frac{\hbar^2}{2m}\nabla^2\Psi + m\Phi_{SG}\Psi\ +\frac{\lambda\hbar^3}{8m^3c}|\Psi|^2\Psi, \label{eq:schrodinger} \\
    \nabla^2\Phi_{SG} &=& 4\pi G|\Psi|^2\ . \label{eq:poisson} 
\end{eqnarray}
Here $\Phi_{SG}$ is the self-gravitational potential of the scalar field configuration. We have defined $\Psi$ such that $\rho(\Vec{r}) = |\Psi(\Vec{r})|^2$ corresponds to the mass density. In the absence of self-interactions ($\lambda = 0$), the system reduces to the well-known Schr\"{o}dinger-Poisson system. 
Before proceeding, we'd like to address a few possible conceptual concerns:

\begin{enumerate}[(a)]
    \item Since the above equations will be used to model DM at galactic scales, factors of $\hbar$ in eq. (\ref{eq:schrodinger}) suggest that there are quantum mechanical effects at galactic length scales. This should not be surprising - the reason atoms are as large as they are is that they are quantum mechanical systems (hence $\hbar$  important) which involve motion of electrons of mass $m_e$ under Coulomb interactions determined by $e$ - the only length scale which can be formed out of these quantities is Bohr radius. Thus, the smallness of atoms (compared to human scales) is not just due to quantum mechanics (i.e. $\hbar$) but also due to the values of $m_e$ and $e$. For ultra-light dark matter, the length scale formed from $m$, $\hbar$ and typical velocity of a particle under the gravitational influence of a galaxy is of Galactic scale. This fact, that quantum mechanical effects are important, is useful in thinking about phenomena such as tunneling which are dealt with in this paper.
    \item In the model of DM we are interested in, since DM particles are Bosons, a lot of them can be in the same (single particle) quantum state - we can thus form coherent states which are well described by classical field theory just like classical electromagnetic waves can be understood in terms of photons. The description of a classical electromagnetic wave doesn't require us to pay attention to the microscopic description in terms of photons. In the same way, under the circumstances of the current problem, we can also describe DM in a galaxy completely in terms of classical field equations.
    \item In particular, in classical Klein-Gordon field theory, $\hbar$ and $m$ don't play any role: the only dimensionful quantities are $c$ and the length scale $L$ in Klein-Gordon equation $( - \partial^2 + L^{-2} ) \varphi = 0$. Working in the convention in which $\varphi$ has dimensions $M^{\frac{1}{2}} T^{-\frac{1}{2}}$, when we introduce a self-interaction term $g_4 \varphi^3$, $g_4$ has dimensions $L^{-2} M^{-1} T^{1}$, we can then express the dimensions of any quantity (e.g. $G$) in terms of $L, c ~{\rm and}~ g_4$. Alternatively, if we wish to include the effect of gravity, we can express all quantities (e.g. $g_4$) in terms of $L, c ~{\rm and}~ G$. In particular, $\Psi$ can be defined from $\varphi$ by an equation which doesn't involve $\hbar$ and $m$ i.e. 
    $\varphi = \sqrt{\frac{c}{2}} \cdot L \cdot \left(e^{-i(ct/L)}\Psi + c.c.\right)$ and
    we can obtain a version of eq. (\ref{eq:schrodinger}) which doesn't involve factors of $\hbar$ and $m$ i.e.
    \begin{equation}
        \frac{iL}{c}  \frac{\partial \Psi}{\partial t} = - \frac{L^2}{2} \nabla^2 \Psi + \frac{\Phi_{SG} \Psi}{c^2} + \frac{L^2 G}{8 c^2} \lambda |\Psi|^2\Psi \; , 
    \end{equation}
    which is a purely macroscopic classical description involving $L, c ~{\rm and}~ G$ (note that $\lambda$ is dimensionless). 
\end{enumerate}

One can thus work with either the microscopic quantum mechanical picture (which involves $\hbar$ and $m$) or macroscopic classical picture involving $L$.

\subsection{Stationary states}\label{sec:stationary_states}

Often, solutions of equations (\ref{eq:schrodinger}) and (\ref{eq:poisson}) are used to model cores of dark matter halos. For this, one looks for a solution (called soliton) which
\begin{inparaenum}[(a)]
  \item is regular everywhere,
  \item has spherical symmetry $\Psi(\Vec{r},t) = \Psi(r,t)$ and $\Phi_{SG}(\Vec{r},t) = \Phi_{SG}(r,t)$,  
  \item is nodeless,
  \item is stationary in the sense that $\Psi(r,t) = e^{-i\gamma t/\hbar}~\phi(r)$ (where $\gamma$ is real and time independent and hence $\phi(r)$ can be taken to be real) and $\Phi_{SG}(r,t) = \Phi_{SG}(r)$, and,
  \item is localized in the sense of being square integrable, i.e. $\int_0^\infty 4\pi r^2|\Psi|^2dr$ is finite.
\end{inparaenum}

Given appropriate boundary conditions (see eq.~(\ref{eq:BCs})), one can then solve for the eigenvalue $\gamma$ to obtain a unique solution $\phi(r)$. The reality of the eigenvalue $\gamma$ is closely connected to the property (e) mentioned above. Time-dependence for a stationary state is purely contained within the phase (with a time independent $\gamma$), implying that the density $\rho(r) = |\Psi|^2 = \phi^2$ is time independent.
In the present paper, we will use such stationary solutions when we wish to recover the case with no tidal effects.

\subsection{Tidal effects}\label{sec:tidal_effects}

Consider two point particles near each other falling freely under the gravity of an external, far away object. The acceleration of the relative position, ${\bf r}'$  of the second particle w.r.t. the first particle, is determined by the Hessian matrix of the gravitational potential due to the far away object. Given this acceleration, one can find what is called tidal potential, which is the potential corresponding to the acceleration determined by solving ${\bf a} = - {\bf \nabla} \Phi_{\rm tidal}$ for $\Phi_{\rm tidal}$ i.e. it is given by
\begin{equation}
    \Phi_{\rm tidal} ({\bf r}) -   \Phi_{\rm tidal} ({\bf r}_0) = - \int_{{\bf r}_0}^{\bf r} {\bf a}({\bf r}') \cdot d {\bf r}' \; .
\end{equation}
In order to define this tidal potential unambiguously, we choose ${\bf r}_0$ to be the location of the first particle, which is the origin ${\bf 0}$ and define $\Phi_{\rm tidal} ({\bf r}_0) = 0$. 

As an example, if the far away object has mass $M$ and is located at $(0, 0, d)$ and if the first particle is at the origin and the second particle is at ${\bf r}$, the acceleration of the separation vector ${\bf r}$ is then given by, 
\begin{equation}\label{eq:tidal_acceleration}
    a^i = -\delta^{ij}\left(\frac{\partial^2\Phi}{\partial x^j \partial x^k}\right)_{\bf 0}r^k = -\frac{GM}{d^3}\left(x\hat{x} + y\hat{y} - 2z\hat{z}\right)\ .
\end{equation}
Defining a tidal potential that leads to the acceleration in eq.~(\ref{eq:tidal_acceleration}) such that $\Phi_\text{tidal}({\bf 0}) = 0$, which is given by

\begin{equation}
    \Phi_\text{tidal} = -\frac{GM}{2d^3}\left(2z^2 - x^2 -y^2\right)\ .
\end{equation}
Note that along the line joining the first particle and far away object, this will go as $\Phi_{\rm tidal} \propto - \omega^2 r^2$, where, $\omega^2 = \frac{GM}{d^3}$ from Kepler's third law.

The above formalism can be applied to a satellite dwarf galaxy orbiting the centre of a host halo at a distance $a$ from halo's centre. In particular for Fuzzy Dark Matter (ULDM without self interactions), where the density profile is given by stationary solutions of the time-independent SP equations, the authors of \cite{Hui:2016ltb, Du:2018qor} employ a spherically symmetric tidal potential due to the halo $\Phi_H$

\begin{equation}\label{eq:tidal_potential}
    \Phi_H = -\frac{3}{2} m\omega^2r^2\ ,
\end{equation}
as an external potential in the SP system to account for the host halo's tidal effects (note the fact that $\Phi_H \propto -\omega^2 r^2$). Here $\omega = \sqrt{\frac{GM_{enc}}{a^3}}$ is the orbital frequency of the satellite and $M_{enc}$ is the mass enclosed within a sphere of radius $a$. 

Recently, \cite{Hertzberg:2022vhk} has provided an analytic derivation of the time-independent Schrödinger-Poisson equations in the presence of a tidal potential, arriving at eq.~(\ref{eq:tidal_potential}) for the $\ell = 0$ mode of the SP system. They show that higher modes of the wave-function will be very small as compared to the $\ell = 0$ mode and that the numerical factor $3/2$ is valid only when $M_{enc} \approx M_{tot}$ for the host halo, i.e. the satellite is orbiting far enough away from the centre of the halo. In this paper, since our aim is to study the effects of self-interactions alone, we shall fix the numerical factor to $3/2$. 

Following the discussion in section~\ref{sec:tidal_effects}, in the presence of a tidal potential, as well as self-interactions, the time-independent spherically symmetric Gross-Pitaevskii-Poisson system can now be written as 

\begin{eqnarray}
   \gamma\phi &=& -\frac{\hbar^2}{2m}\nabla_r^2\phi + \left(m\Phi_{SG} - \frac{3}{2}m\omega^2r^2 + \frac{\lambda\hbar^3}{8m^3c}|\phi|^2\right)\phi\ ,\label{eq:mod_GP} \\ 
   \nabla_r^2\Phi_{SG} &=& 4\pi G|\phi|^2\ , \label{eq:mod_poisson}
\end{eqnarray}
where $\nabla_r^2 = \frac{\mathrm{d}^2}{\mathrm{d}r^2} + \frac{1}{r}\frac{\mathrm{d}}{\mathrm{d}r}$. Here we have assumed that similar to the no self-interactions case, tidal effects will only lead to an additional $\Phi_H$ term in the Gross-Pitaevskii (GP) equation. It is worth noting that $\Phi_H$ is included as an external potential while $\Phi_{SG}$ must simultaneously satisfy the Poisson equation. One can write eq.~(\ref{eq:mod_GP}) such that it resembles a time-independent Schrödinger equation where one can define an effective potential, $V_\text{eff}$ using the terms in the parenthesis on the RHS, 

\begin{equation}\label{eq:effective_potential}
    V_{\text{eff}} = \Phi_{SG} + \Phi_H + \Phi_{SI}\ .
\end{equation}
Here $\Phi_{SI} = \frac{\lambda\hbar^3}{8m^3c}|\phi|^2$. This effective potential will act as the potential barrier for a tunnelling problem. For intuition, we have plotted the effective potential in its dimensionless form (defined in section~\ref{sec:numerical_procedure}) in the absence (figure~\ref{fig:barrier}) and presence (figure~\ref{fig:SI_barrier}) of self-interactions. Note that in both panels, the curves are obtained after solving the modified GPP system numerically.

Our goal is to solve Eqs (\ref{eq:mod_GP}), (\ref{eq:mod_poisson}) and find how the scalar self-coupling $\lambda$ affects the rate at which DM tunnels out of the satellite galaxy.

\subsection{Quasi-stationary states}\label{sec:quasi-stationary}

If we think of tunneling as merely causing a slow leakage of DM from an otherwise stationary solution, we can employ the formalism of quasi-stationary states (see section \textsection 134 of \cite{Landau:2004qm} and the last chapter of \cite{Perelomov:1998qmst}). In this approach, one does two (closely related) things: 

\begin{itemize}
    \item Allow the ``energy" eigenvalue $\gamma$ in eq (\ref{eq:mod_GP})  to be complex i.e. $\gamma = \gamma_R + i\gamma_I$
    \item Look for solutions which satisfy an outgoing wave boundary condition.
\end{itemize}
Solutions with $\gamma_I\neq 0$ satisfying the outgoing boundary conditions are known as quasi-stationary states and are often used to model tunnelling of particles through potential barriers in quantum mechanics \cite{Landau:2004qm, Perelomov:1998qmst}.

Note that for $\gamma_I < 0$, the classical field $\Psi$ will have an exponentially decaying part since $e^{-i\gamma t/\hbar} = e^{-i\gamma_R t/\hbar}e^{\gamma_It/\hbar}$ implying that the density is no longer constant in time: $\rho = |\Psi|^2 = e^{2\gamma_It/\hbar}|\phi|^2$, i.e. we are no longer dealing with stationary states. In fact, since $\gamma_I < 0$, the density will decrease with time, which can be interpreted as dark matter leaking out through the barrier.

A consequence of tunnelling is that beyond the barrier, the solution will not be zero and will resemble an outgoing wave. Using the WKB approximation one can obtain this asymptotic behaviour at large $r$ where the $ -\frac{3}{2}\omega^2 r^2$ term in eq.~(\ref{eq:mod_GP}) dominates over other terms in the parentheses, to be 
\begin{equation}\label{eq:WKB}
    \lim_{r \to \infty} \phi(r) = \phi_{\text{WKB}}(r) = \frac{A}{r^{3/2}}\exp{\left(i\frac{\sqrt{3}}{2\hbar}m\omega r^2\right)}\ .
\end{equation}
Note that this behaviour at large $r$ will lead to a solution which is no longer square integrable (recall the conditions mentioned in section \ref{sec:stationary_states}) - it is logarithmically divergent at large $r$.

A complex $\gamma$ forces $\phi$ to be complex as well; $\Phi_{SG}$ on the other hand depends on $|\phi|^2$ and hence will always be real. Since $\phi$ can no longer be real and since its real and imaginary parts oscillate at large $r$, where outgoing wave boundary conditions, eq (\ref{eq:WKB}), are applicable (though its absolute value doesn't go to zero at any finite $r$), we can no longer say that we are looking for a nodeless solution in the usual sense of the word.

In summary, among the five conditions mentioned in section \ref{sec:stationary_states}, which hold good for stationary solutions,  only the first two (regularity and spherical symmetry) continue to hold good for quasi-stationary solutions.
We are thus dealing with a solution which is no longer localised i.e. it is non-normalisable. In the parlance of QM, the particle is able to go away to infinity - we no longer have bound states. For such solutions, the eigenvalue $\gamma$ is not restricted to be real and $\int_0^{\infty} |\Psi|^2 r^2 dr$ is time dependent and the wave function doesn't die down fast enough at large $r$ for this to be strictly convergent.

In the next section, we shall solve system in eqs.~(\ref{eq:mod_GP}) and~(\ref{eq:mod_poisson}) as an eigenvalue problem, by searching for the correct values of $\gamma_R$ and $\gamma_I$ such that far from the origin, the solution is matched with the WKB solution.
In appendix \ref{app:approx_appraoch}, we describe an approximate approach which is helpful in developing an intuitive understanding which is especially useful because it is very similar to the approach taken in most standard texts while describing tunneling.

\section{Numerical implementation}\label{sec:numerical_procedure}
To solve the system numerically, we introduce dimensionless variables, denoted by a ` $\hat{}$ ' over quantities, in terms of $m$ and fundamental constants

\begin{eqnarray}
    \hat{\phi} = \frac{\hbar\sqrt{4\pi G}}{mc^2}\phi, \ \ \hat{\Phi} &=& \frac{\Phi}{c^2}, \ \ \hat{\gamma} = \frac{\gamma}{mc^2}, \ \ \hat{t} = \frac{mc^2}{\hbar}t \label{eq:dimensions1} \\  \hat{r} = \frac{mc}{\hbar}r, \ \ \hat{\lambda} &=& \frac{\lambda}{64\pi}\left(\frac{m_{pl}}{m}\right)^2, \ \ \hat{\omega} = \frac{\hbar}{mc^2}\omega\ .\label{eq:dimensions2}
\end{eqnarray}
Here $\hat{r}$ is now in units of the Compton wavelength of the scalar field, while $\hat{t}$ is in the units of `Compton time'. The free parameters of the system now are $\hat{\lambda}$ and $\hat{\omega}$ (note that dimensionless version of the system of equations will not depend on $m$ using the re-definitions above). We now have two sets of initial and boundary conditions depending on the presence or absence of the tidal potential: 

\begin{itemize}
    \item For $\hat{\omega} = 0$, the system admits a stationary solution representing a bound state with the ground state eigenvalue $\hat{\gamma}$. We impose the following boundary conditions\footnote{One can always redefine the self-gravitational potential by adding a constant, $\hat{\Phi}_{SG} \equiv \hat{\Phi}_{SG} + C$ without changing the form of Poisson equation in eq.~(\ref{eq:mod_poisson}). In eq.~(\ref{eq:mod_GP}), one can then define a new $\hat{\gamma} \equiv \hat{\gamma} - C$, leaving the form of the system of equations unchanged.}:
    \begin{equation}\label{eq:BCs}
        \hat{\Phi}_{SG}(0) = 0,\ \hat{\phi}'(0) = \hat{\Phi}_{SG}(0) = 0,\ \hat{\phi}(\infty) = 0\ . 
    \end{equation}
    We begin by setting $\hat{\phi}(0) = 1$, which will uniquely determine the ground state eigenvalue $\hat{\gamma}$. Numerically we will impose the boundary conditions at infinity by imposing them at a large value of $\hat{r}$ i.e. $\hat{r}_m$. One can then find a unique real $\hat{\gamma}$ that will satisfy the above boundary conditions, and the corresponding solution $\hat{\phi}$ is called a soliton. 

    \item For $\hat{\omega} > 0$, the boundary conditions are different. Since the solution is not a stationary state, $\hat{\gamma} = \hat{\gamma}_R + i\hat{\gamma}_I$. The first three conditions in eq.~(\ref{eq:BCs}) remain the same as the previous case, while for behaviour at large $\hat{r}\rightarrow \infty$, we now have outgoing waves. We impose the BCs at infinity by choosing a large $\hat{r}_m$, and matching the numerical solution as well as its first derivative with the asymptotic solution in eq.~(\ref{eq:WKB}), i.e., $\hat{\phi}(\hat{r}_m) = \hat{\phi}_\text{WKB}(\hat{r}_m)$ and $\hat{\phi}'(\hat{r}_m) = \hat{\phi}_\text{WKB}'(\hat{r}_m)$ (note that this must hold for both real and imaginary parts of $\hat{\phi}$). Similar to the stationary state case, we set $\text{Re}(\hat{\phi}(0)) = 1\ , \text{Im}(\hat{\phi}(0)) = 0$. We look for solutions with $\hat{\gamma}_I < 0$, all of these conditions will uniquely determine the values of $\hat{\gamma}_R$ and $\hat{\gamma}_I$.\footnote{Note that for every solution that satisfies the boundary condition in eq.~(\ref{eq:WKB}), we ensure that the density has negligible `wiggles' and $|\hat{\phi}|^2/|\hat{\phi}_{\text{WKB}}|^2 \propto r^0$ up to a certain accuracy (we require the slope of the above function to be $< 0.05$) around $\hat{r}_m$ to ensure appropriate behaviour.} While the solution obtained here is not a stationary state, we continue to call it a soliton for the rest of the paper.
\end{itemize}

\begin{figure}[ht]
    \centering
    \includegraphics[width = 0.7\textwidth]{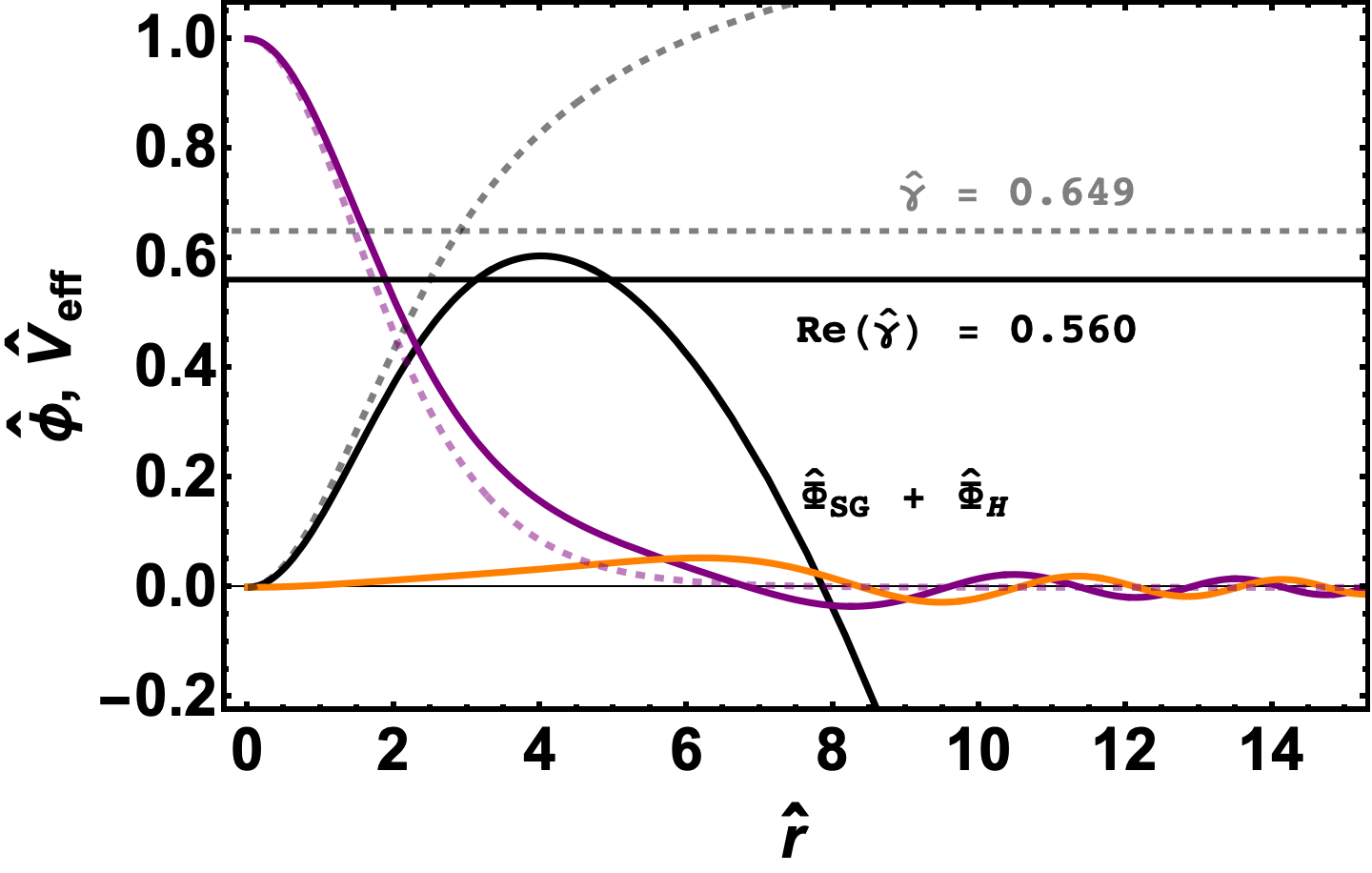}
    \caption{A numerical solution with (solid curves) and without (dashed curves) tidal potential is shown. For $\hat{\omega} = 0.12$, the real part of the quasi-stationary solution is shown by the purple solid curve while the imaginary is shown by the orange solid curve. Note that the real part of the eigenvalue for the quasi-stationary $\hat{\gamma}_R$ is smaller than $\hat{\gamma}$ for the stationary state. Also note the presence of a local maximum in the effective potential when a tidal potential is included.}
    \label{fig:example_sol}
\end{figure}

An example solution found using the shooting method \cite{Giordano:2006ncp} for $\hat{\omega} = 0$ and $\hat{\omega} = 0.12$ with $\hat{\lambda} = 0$ is shown in figure~\ref{fig:example_sol}.

\subsection{Tunnelling rate and lifetime}\label{sec:decay_rates}

Once we have a solution, we have a density profile $\hat{\rho}(\hat{r}) = |\hat{\phi}(\hat{r})|^2$ and can define the mass enclosed in a sphere of radius $\hat{r}$ as

\begin{equation}\label{eq:soliton_mass}
    \hat{M}(\hat{t}, \hat{r}) = \frac{G m}{\hbar c}M(t, r)= \int_0^{\hat{r}}|\hat{\Psi}(\hat{t}, \hat{r}')|^2 \hat{r}'^2 d\hat{r}' = e^{2\hat{\gamma}_I t}\left[\int_0^{\hat{r}}|\hat{\phi}(\hat{r}')|^2 \hat{r}'^2 d\hat{r}'\right]\ , 
\end{equation}
where we have used eqs.~(\ref{eq:dimensions1}) and~(\ref{eq:dimensions2}) and $\hat{\gamma}_I \leq 0$. 

It is important to note that for $\hat{\omega} = 0$, $\hat{\gamma}_I = 0$ and hence mass of the soliton will remain constant over time. On the other hand, for $\hat{\omega} > 0$, from eq.~(\ref{eq:soliton_mass}) and the fact that $\hat{\gamma}_I < 0$ the soliton will lose its mass over time exponentially. We can now define a (dimensionless) decay rate as

\begin{equation}
  \hat{\Gamma} \equiv \frac{-\left(\mathrm{d}\hat{M}/\mathrm{d}\hat{t}\right)}{\hat{M}} = -2\hat{\gamma}_{I}\ ,
\end{equation}
where it is to be noted that $\hat{\gamma}_I$ and hence $\hat{\Gamma}$ are time-independent. Note that this dimensionless decay rate is related to the dimensionful decay rate by the relation $\hat{\Gamma} = \left(\hbar/mc^2\right)\Gamma$. Lifetime of the soliton is just the inverse of the decay rate (or, tunnelling rate) i.e., $\tau \equiv \Gamma^{-1}$.

\subsubsection{Core radius and core mass}

In order to calculate any measure of mass of the soliton, one needs to evaluate the integral inside the square brackets in eq.~(\ref{eq:soliton_mass}). Note that one cannot define a total mass (found by replacing the upper limit by $\hat{r}\rightarrow \infty$) for quasi-stationary states, since boundary conditions in eq.~(\ref{eq:WKB}) imply that at large $\hat{r}$, $|\phi|^2\propto r^{-3}$ making the integral in eq.~(\ref{eq:soliton_mass}) log divergent. We can however define some measure of mass by first defining the size of the soliton as the radius at which the density becomes half its central value (also called core radius $\hat{R}_c$). This can then be used to define a soliton mass as the mass contained within core radius $\hat{M}_c = \hat{M}(\hat{R}_c)$ (also called core mass).

\subsection{Scaling relations}\label{sec:scaling_relations}

For a fixed $m$, $\hat{\lambda}$, $\hat{\omega}$, and $\hat{\phi}(0)$, the ground state eigenvalue $\hat{\gamma}$ and hence the density profile $|\hat{\phi}(\hat{r})|^2$ is uniquely determined by the boundary conditions. The density profile in-turn also fixes soliton mass and size as defined in eq.~(\ref{eq:soliton_mass}). To describe a different soliton, one can either numerically solve the system again for a different value of $\hat{\phi}(0)$ or one can scale every variable in a manner that preserves the form of the system of equations (see e.g. \cite{Chakrabarti:2022owq} for details). 

For instance, suppose we scale the length associated with the system by some amount $r\rightarrow sr$. To ensure that all terms in eqs.~(\ref{eq:mod_GP}) and~(\ref{eq:mod_poisson}) transform identically, we must simultaneously scale all variables in the following manner: 

\begin{equation}\label{eq:scaling}
    \{r, \phi, \Phi, \gamma, \omega, \lambda, t\} \rightarrow \{sr, s^{-2}\phi, s^{-2}\Phi, s^{-2}\gamma, s^{-2}\omega, s^2\lambda, s^2t\}\ .
\end{equation}
The above property is also called scaling symmetry \cite{Guzman:2006yc, Davies:2019wgi, Li:2020ryg} (see also \cite{Chakrabarti:2022owq}), and is able to characterize an entire family of solitons with different masses, sizes and central densities from a single numerical solution found for a fixed $\hat{\phi}(0)$. For our numerical solutions, we fix $\hat{\phi}(0) = 1$, which from our normalization scheme in eqs.~(\ref{eq:dimensions1}) and~(\ref{eq:dimensions2}) describes an unrealistic solution with $\rho_c \sim 10^{14}\ \text{M}_\odot/\text{pc}^3$. Hence, $s \gg 1$ is required to scale central density down to a physical solution (see \cite{Chakrabarti:2022owq} for details). For the remainder of the paper, we shall use the subscript `\text{ini}' to denote quantities with $s = 1$, i.e. unscaled and use `\text{fin}' for scaled quantities. 
Note that from eqs.~(\ref{eq:soliton_mass}) and~(\ref{eq:scaling}), enclosed mass scales as $M \rightarrow s^{-1}M$, which as an example of our notation, can be written as $\hat{M}_{\text{fin}} = s^{-1}\hat{M}_{\text{ini}}$. Similarly, since $\Gamma$ has the same dimensions as $\gamma_I$, it also scales as $\Gamma \rightarrow s^{-2}\Gamma$. 

Now that we have set up the machinery of solving the system, we shall look at how changing $\hat{\omega}$ and $\hat{\lambda}$ alters decay rates and corresponding lifetimes of satellite galaxies. It is worth noting the range of parameters we consider throughout the reminder of the paper: ${\hat \omega}_{\text{ini}} \in [0.03, 0.155]$ and ${\hat \lambda}_{\text{ini}} \in [-0.4,0.2]$ while $m = 10^{-22}\ \text{eV}$ is fixed.

\subsection{Decay rates vs. $\rho_c/\rho_H$}\label{sec:comparison}

The results of recent work \cite{Hui:2016ltb, Du:2018qor, Hertzberg:2022vhk} suggests that for FDM, the rate of decay or mass loss depends only on the ratio of the central density of the satellite galaxy $\rho_c$ and the average density of the host halo within the orbital radius $\rho_H$. In-fact, they obtain a fitting function from numerical solutions that suggests that the decay rate exponentially decreases with an increasing $\rho_c/\rho_H$: $\Gamma \propto e^{-\rho_c/\rho_H}$. 

In order to compare our solutions with previous work, we first note that for a fixed $\hat{\lambda}_\text{ini}$ a different $\hat{\omega}_\text{ini}$ will lead to a different solution. Now using scaling relations in eq.~(\ref{eq:scaling}), it is straightforward to see that $\hat{\rho}_{0, \text{fin}}/\hat{\omega}_{\text{fin}}^2 = \hat{\rho}_{0, \text{ini}}/\hat{\omega}_{\text{ini}}^2$ where $\hat{\rho}_{0,\text{ini}}$ and $\hat{\rho}_{0,\text{ini}}$ are the unscaled and scaled values of density at the origin, i.e. the central density of the satellite galaxy. If the satellite is orbiting the halo centre at a radius $a$ with orbital period $T$, one can define the average density of the host halo in a sphere of radius $a$ as $\rho_H = \frac{3M_{enc}}{4\pi a^3} = \frac{3\omega^2}{4\pi G}$ (where we have used Kepler's third law and $\omega = 2\pi/T$ is the scaled orbital frequency of the satellite). Denoting the scaled central density as $\rho_c \equiv \rho_{0,\text{fin}}$ one can write the ratio of the scaled dimensionful central density and average halo density as, 

\begin{equation}\label{eq:density_ratio}
    \frac{\rho_c}{\rho_H} = \frac{\hat{\rho}_{0,\text{fin}}}{3\hat{\omega}_\text{fin}^2} = \frac{1}{3}\frac{s^{-4}\hat{\rho}_{0,\text{ini}}}{s^{-4}\hat{\omega}_\text{ini}^2} = \frac{1}{3\hat{\omega}_\text{ini}^2}\ ,
\end{equation}
where we have $\hat{\rho}_{0,\text{ini}} = |\hat{\phi}(0)|^2 = 1$ is fixed in our numerical solutions. Hence, changing $\hat{\omega}_\text{ini}$ implies changing the ratio of central density and average halo density.

Now, for different values of $\hat{\omega}_\text{ini}$, one can always choose scale values $s$ such that either $\hat{\rho}_{0,\text{fin}}$ or $\hat{\omega}_\text{fin}$ remain fixed. Hence a changing $\rho_c/\rho_H$ can imply either changing central density or a changing the orbital frequency. A higher $\rho_c/\rho_H$ can therefore imply that the satellite has a higher central density or a smaller orbital frequency (i.e. a larger orbital radius if $M_{enc}$ is fixed). 

\subsubsection{Recovering the case of negligible DM self-interactions}

We obtain decay rates for different values of $\hat{\omega}_{\text{ini}}$ by solving the system of equations in eqs.~(\ref{eq:mod_GP}) and~(\ref{eq:mod_poisson}) numerically. We then plot decay rates $\Gamma/(\omega/2\pi)$ versus the ratio of central density of the satellite and the average density of the host halo $\rho_c/\rho_H$ in figure~\ref{fig:decay_vs_ratio} (blue curve). We then compare our solutions with the fitting function obtained by \cite{Hertzberg:2022vhk} (black curve) and find that the decay rates obtained by us agree well with it. 

In the context of potential barriers, one can understand why a larger $\hat{\omega}_\text{ini}$ leads to a larger decay rate using potential barriers from figure~\ref{fig:approx_pot_barrier} (approximate potential barriers). It is clear that a larger $\hat{\omega}_\text{ini}$ implies a narrower potential barrier which in the tunnelling picture should make it easier for DM to tunnel out, leading to a shorter lifetime, while a smaller $\hat{\omega}_\text{ini}$ implies a wider barrier and hence a longer lifetime.

\subsubsection{Core mass and lifetime}

Now, recall from section~\ref{sec:decay_rates} that we can define a core radius $\hat{R}_c$ as the distance at which the density becomes $1/2$ its central value, and hence define the mass of the core as $\hat{M}_{\text{ini}}\equiv\hat{M} (\hat{R}_c)$ using eq.~(\ref{eq:soliton_mass}). Using scaling relations in eq.~(\ref{eq:scaling}), one can write the central density in terms of scaled core mass $M_c \equiv M_{\text{fin}} = s^{-1}\frac{\hbar c}{Gm}\hat{M}_\text{ini}$ as, 

\begin{equation}\label{eq:mass_density}
   M_c^4  \propto \frac{\hbar^6}{G^3m^6}\rho_c\ .
\end{equation}
Here, for a fixed $\omega_{\text{fin}}$ if the scaled core mass $M_c$ is larger, then $\rho_c/\rho_H$ is also larger and consequently the lifetime of the soliton is longer. In other words, heavier satellites survive for longer than lighter satellites. It was suggested in \cite{Hui:2016ltb, Hertzberg:2022vhk} that in general for a typical dwarf satellite, $\rho_c/\rho_H \gtrsim 70$ is required for it to survive till present day in orbit around a host halo.

\subsubsection{The impact of self-interactions of DM: a first look}

\begin{figure}[ht]
    \centering
    \includegraphics[width = 0.7\textwidth]{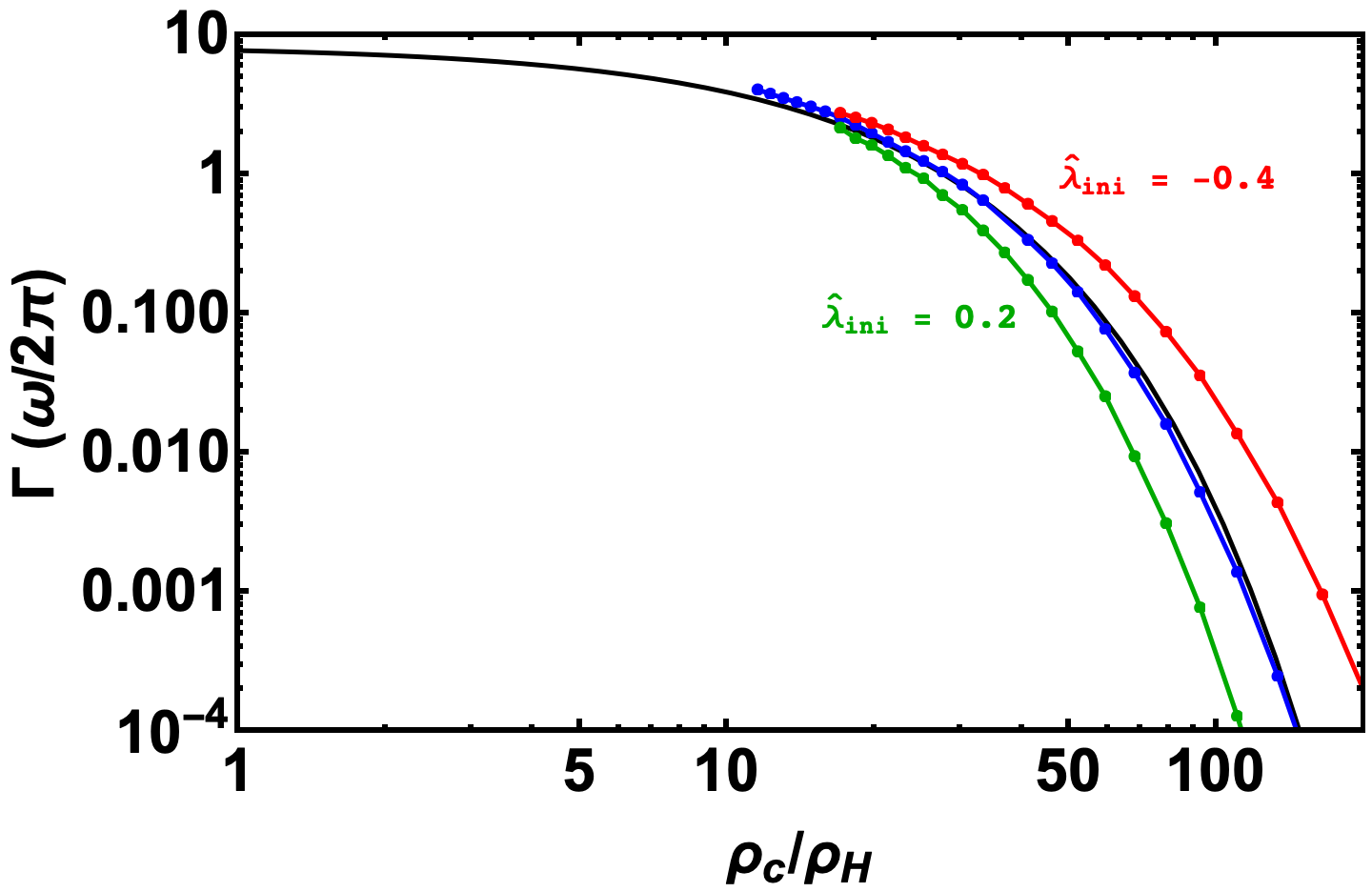}
    \caption{We have plotted decay rates (in the units of $T^{-1} = \omega/2\pi$) vs. the ratio of central density of the satellite and average density in the host halo. The solid black line is the fitting function obtained in \cite{Hertzberg:2022vhk} for $\lambda = 0$. The blue curve are the decay rates obtained using our numerical implementation while red and green curves are decay rates for a fixed unscaled dimensionless attractive and repulsive self-interaction strength $\hat{\lambda}_\text{ini} = -0.4, 0.2$ respectively.}
    \label{fig:decay_vs_ratio}
\end{figure}

This procedure can be repeated in the presence of self-interactions: for two values, $\hat{\lambda}_\text{ini} = -0.4$, and $\hat{\lambda}_\text{ini} = 0.2$, we show the corresponding decay rates in figure~\ref{fig:decay_vs_ratio} as the red and green curves respectively. 
It is also worth noting that for the case of attractive self-interactions, we only look at solutions corresponding to $\hat{\lambda}_{ini} > -0.4$. This is because in the absence of tidal effects, $\hat{\lambda}_{ini} < -0.4$ corresponds to the unstable branch of solutions (or the so-called non-gravitational regime) \cite{Dave:2023wjq, Chavanis:2011mrr}. 

We see that just like the case of no interactions, a larger $\hat{\omega}_\text{ini}$ leads to a larger decay rate and a smaller lifetime. However, we also see that for a fixed $\hat{\omega}_\text{ini}$ attractive self-interactions appear to have a higher decay rate than repulsive self-interactions. This suggests that attractive self-interactions lead to a satellite galaxy surviving for a shorter time before losing most of its mass to tidal effects while repulsive self-interactions allow the satellite to survive for longer. This result bears a closer inspection, which we shall carry out in section~\ref{sec:fixed_rhoC}.

To understand exactly how these solutions translate to describing realistic objects, we must employ scaling relations, as we shall see in the next section~\ref{sec:fixed_rhoC}.

\section{Impact of self-interactions}\label{sec:SI_impact}
In this section we shall look at how one can use scaling relations to model realistic objects and analyse the impact of self-interactions in two different ways. In section~\ref{sec:fixed_rhoC} we shall look at the case where the scaled central density is fixed along with $\omega_\text{fin}$, and allow $\lambda$ to vary. Next, in section~\ref{sec:fixed_Mc} we shall look at objects with fixed mass (in this case, core mass) $M_\text{fin}$ and orbital frequency $\omega_\text{fin}$ and see how self-interactions alter corresponding life-times. Finally in section~\ref{sec:saving_ULDM} we shall look under what conditions will self-interactions allow objects to survive that are ruled out for the $\lambda = 0$ case.

Recall that the numerical solutions we obtain are for $s = 1$. However, to model realistic objects we need to obtain an appropriate value of scale which at least satisfies $s \gtrsim 10$ (this describes the `trustworthy regime' where the effects of general relativity will not be important \cite{Chakrabarti:2022owq}). 

We can estimate the order of magnitude of $s$ required by considering the typical orbital period of a satellite to be $T = 2\pi/\omega \sim \mathcal{O}(1)\ \text{Gyr}$ (For instance, $T \approx 1.6\ \text{Gyr}$ and $T \approx 3.9\ \text{Gyr}$ for UMi and Fornax respectively \cite{Hertzberg:2022vhk}). However, for $m = 10^{-22}\ \text{eV}$, $\hat{\omega}_{\text{ini}} \sim 0.05$ implies an unscaled time period of $T_{\text{ini}} = 2\pi/\omega_{\text{ini}} \sim \mathcal{O}(10)\ \text{yr}$ which is far smaller than what is observed. The required scale value can then be obtained from eq.~(\ref{eq:scaling}) as $s \sim \mathcal{O}(10^4)$. Additionally, for an estimate of the central density or the core mass of a satellite, obtaining the correct $s$ that scales $\hat{\omega}_\text{ini}$ and $\hat{M}_\text{ini}$ (or $\hat{\rho}_{c, \text{ini}}$) to required values is more involved, as we shall see in the next sections. 


\subsection{Objects with known central density and orbital period}\label{sec:fixed_rhoC}

Suppose the known quantities for a satellite galaxy are the central DM density and the orbital period around the host halo at the centre. We denote these as $\rho_{c,\text{fin}}$ and $T_\text{fin} = 2\pi/\omega_{\text{fin}}$ respectively. In our numerical solutions, the central density without scaling is fixed since $\hat{\rho}_{c,\text{ini}} = |\hat{\phi}(0)|^2 = 1$ for all values of $\hat{\omega}_\text{ini}$ and $\hat{\lambda}_\text{ini}$. 

\subsubsection{The procedure}

To find the value of scale $s$ that satisfies observed values $\rho_{c, \text{fin}}$ and $\omega_\text{fin}$, we employ the following procedure:

\begin{enumerate}
    \item Fix dimensionless quantities $\hat{\rho}_{c,\text{fin}}$ and $\hat{\omega}_{\text{fin}}$ using the observed values.
    
    \item Using scaling relations in eq.~(\ref{eq:scaling}), obtain scale value $s = (\hat{\rho}_{c, \text{fin}})^{-1/4}$ (recall that $\hat{\rho}_{c,\text{ini}} = 1$). 

    \item Using $s$ and eq.~(\ref{eq:scaling}), obtain the unscaled orbital frequency such that $\hat{\omega}_{\text{ini}} = s^2\hat{\omega}_{\text{fin}}$ is satisfied. 

    \item For a fixed $\hat{\lambda}_{\text{ini}}$ and the obtained $\hat{\omega}_{\text{ini}}$ solve the dimensionless form of eqs.~(\ref{eq:mod_GP}) and~(\ref{eq:mod_poisson}) numerically to obtain the unscaled decay rate $\hat{\Gamma}_{\text{ini}} = 2|\hat{\gamma}_I|$. Scale the decay rate using $s$ to obtain the lifetime of the soliton $\tau = s^2\frac{\hbar}{mc^2}\hat{\Gamma}_\text{ini}^{-1}$. 
\end{enumerate}

\begin{figure}[ht]
    \centering
    \includegraphics[width = 0.7\textwidth]{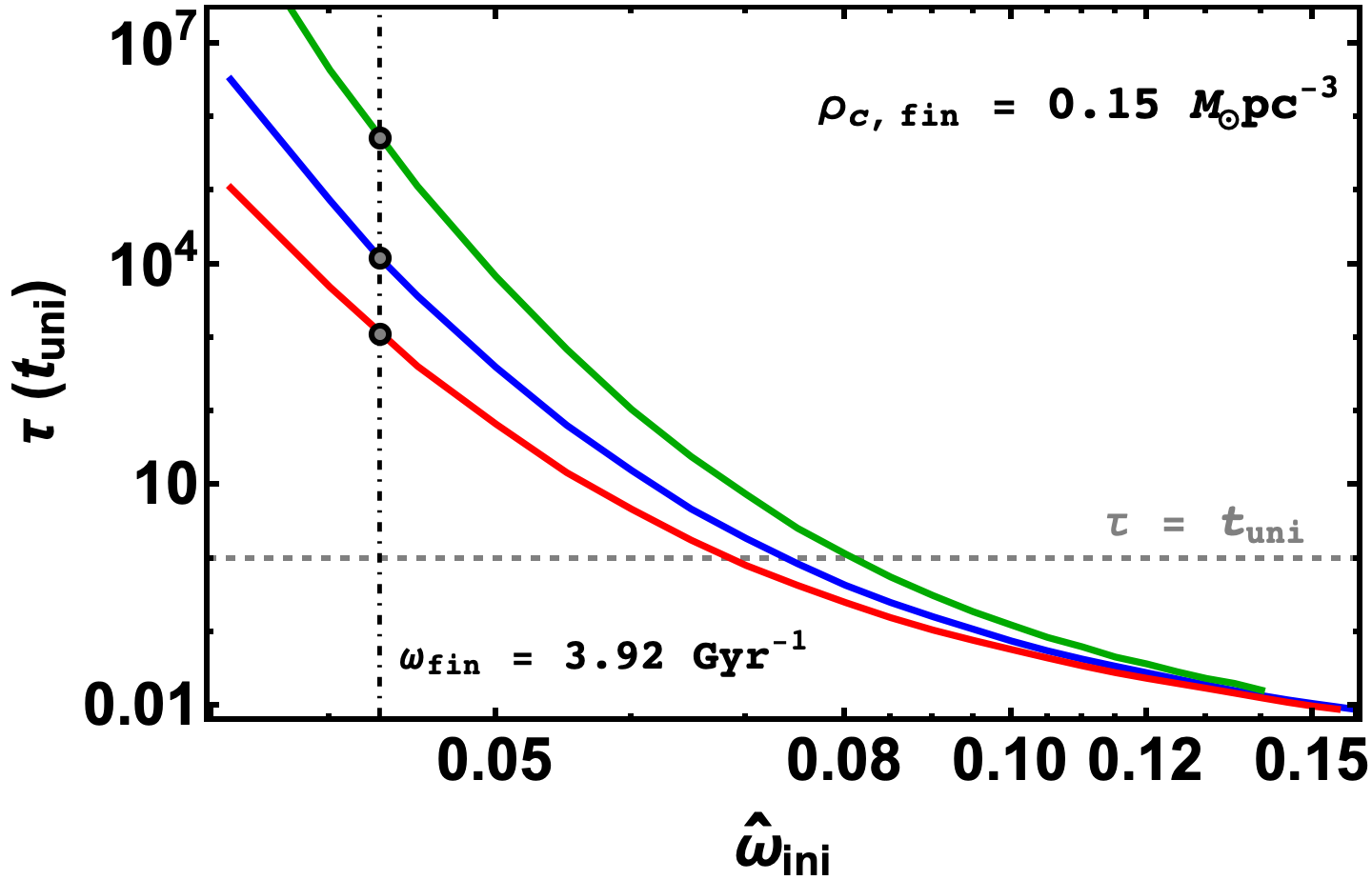}
    \caption{Scaled lifetimes for $\hat{\lambda}_\text{ini} = -0.2, 0, 0.2$ are shown by the red, blue and green curves respectively. Here $s = 7225$ is taken to satisfy a scaled $\rho = 0.15\ \text{M}_\odot\text{pc}^{-3}$. The black dots correspond to lifetimes when the scaled time period is $2\pi/\omega_{\text{fin}} = 1.6\ \text{Gyr}$, i.e. time period of the UMi dwarf galaxy.}
    \label{fig:lifetime_fixed_rho}
\end{figure}

\subsubsection{The absence and presence of self-interactions}
 
To demonstrate the above procedure, we take an example of the Ursa Minor (UMi) spheroidal dwarf with central density $\rho_c = 0.15\ \text{M}_\odot \text{pc}^{-3}$ and $T = 1.6\ \text{Gyr}$ \cite{Hertzberg:2022vhk}. The scale $s$ and $\hat{\omega}_{\text{ini}}$ that satisfy observed values are $s = 7725$ and $\hat{\omega}_{\text{ini}} = 0.0427$. We work initially in the FDM regime, where $\hat{\lambda}_\text{ini} = 0$. The eigenvalue is found to be $\hat{\gamma} = 0.639 - i(3.19\times 10^{-8})$. The scaled decay rate $\Gamma$ and lifetime $\tau$ are then,

\begin{equation}
    \Gamma \approx 0.8\times 10^{-4}\ \text{t}_{\text{uni}}^{-1} \ \  , \ \ \tau \approx 1.23\times 10^4\ \text{t}_{\text{uni}}\ ,
\end{equation}
where, $\text{t}_{\text{uni}} = 13.8\ \text{Gyr}$ is the age of the Universe for the standard $\Lambda$CDM model \cite{Planck:2018vyg}. The above decay rate agrees well with the one obtained by Ref.~\cite{Hertzberg:2022vhk}. The corresponding lifetime of the satellite is far larger than the age of the universe, allowing it to survive until present time. Also note that the scaled core mass of the satellite is $M_c = 9.31\times 10^7\ M_\odot$. 

Now we turn on self-interactions, and choose $\hat{\lambda}_\text{ini} = \pm 0.2$, along with $s = 7225$ and $\hat{\omega}_\text{ini} = 0.0427$. Figure~\ref{fig:lifetime_fixed_rho} plots the lifetime of an object with the above-mentioned properties as a function of $\hat{\omega}_{\text{ini}}$. It is worth noting that the values of $s$ and $\hat{\omega}_\text{ini}$ will be fixed independently of the value of $\hat{\lambda}_\text{ini}$ since $s$ depends on $\hat{\rho}_{0,\text{ini}}$ which is fixed to $1$ for all values of $\hat{\omega}_\text{ini}$ and $\hat{\lambda}_\text{ini}$. The corresponding scaled self-coupling strength is $\lambda = s^2\hat{\lambda}_\text{ini} = \pm 1.41\times 10^{-91}$. The lifetime turns out to be $\tau \approx 5.17\times 10^5\ \text{t}_\text{uni}$ for repulsive self-interactions and $\tau \approx 1.15\times 10^3 \ \text{t}_\text{uni}$ for attractive self-interactions. This corresponds to a larger decay rate for attractive self-interactions while a smaller one for repulsive self-interactions compared to the no self-interactions case which agrees with the what we have seen in figure~\ref{fig:decay_vs_ratio}.

It is worth noting that the above results agree qualitatively with what we obtain from the approximate approach in appendix~\ref{app:approx_appraoch}, where we solve the system without the tidal potential, construct an effective potential with the tidal potential and use WKB approximation inside the barrier to estimate the decay-rate (similar to the $\alpha$-decay problem in quantum mechanics). In the approximate approach, while the numbers do not agree with the numerical solutions, the lifetime does decrease (increase) as one goes from no self-interactions to attractive (repulsive) self-interactions. In particular, note the widening (narrowing) of the barrier for $\lambda > 0$ ($\lambda < 0$) in figure~\ref{fig:wkb_fixed_rhoc} indicating that in the presence of repulsive self-interactions DM is less likely to tunnel out while the presence of attractive self-interactions increases the transmission probability.

\subsubsection{A discrepancy}\label{sec:discrepancy}

Intuitively as well as from simulations \cite{Glennon:2022huu} one expects attractive self-interactions to increase the lifetime of the satellite. In stark contrast, in the results above, we find that attractive self-interactions lead to a smaller lifetime while repulsive self-interactions lead to the satellite galaxy surviving for longer. What exactly is happening here?

\begin{figure}[ht]
    \centering
    \includegraphics[width = 0.65\textwidth]{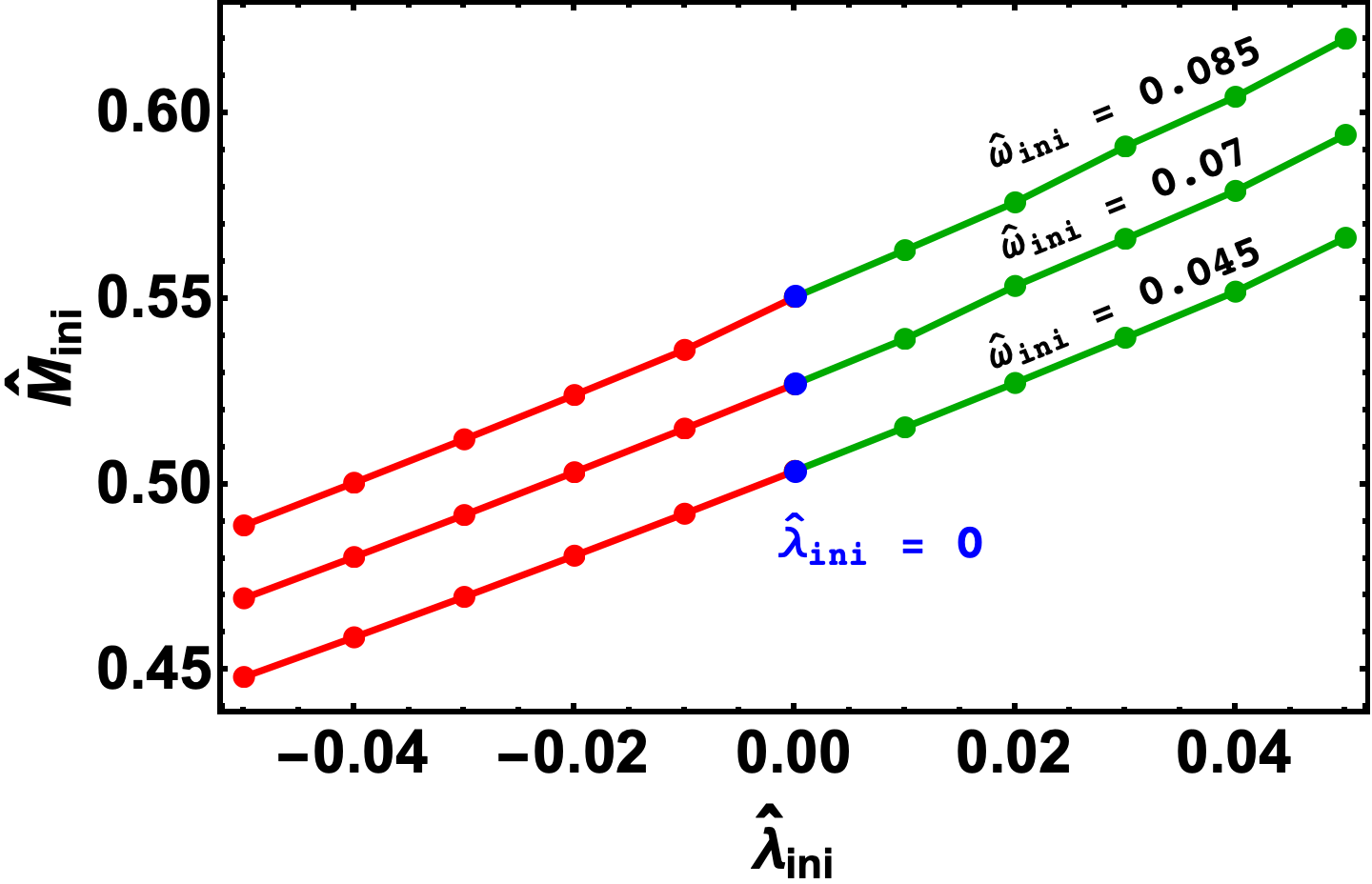}
    \caption{Here we plot the unscaled dimensionless core mass $\hat{M}_\text{ini}$ versus unscaled dimensionless self-interaction strength $\hat{\lambda}_\text{ini}$ for different values of $\hat{\omega}_\text{ini}$. Core mass is defined as the mass contained within a sphere of radius $\hat{R}_{1/2}$, where $\hat{R}_{1/2}$ is the distance at which density becomes half its central value}
    \label{fig:mhat_vs_lambda}
\end{figure}

To understand this discrepancy, we note that for $\hat{\lambda}_{\text{ini}} > 0$ ($\hat{\lambda}_\text{ini} < 0$) any integrated mass $\hat{M}_\text{ini}$ is found to be larger (smaller) than the $\hat{\lambda}_\text{ini} = 0$ case (for a fixed value of $\hat{\omega}_\text{ini}$), as shown in figure~\ref{fig:mhat_vs_lambda}.  

As we have mentioned earlier, for a fixed scaled central density, the scale value $s$ remains the same for different values of $\hat{\lambda}_\text{ini}$. This implies that similar to unscaled core masses, scaled core masses will also be larger for repulsive self-interactions and smaller for attractive self-interactions. In other words, when we fix the scaled central densities for different values of $\hat{\lambda}_\text{ini}$, the scaling relations ensure that we are looking at objects with different scaled soliton masses. Further, as we have seen from eq.~(\ref{eq:mass_density}), a large core mass for a fixed orbital frequency leads to a smaller decay rate and a longer lifetime, and vice-versa. This reasoning also explains the effect of self-interactions on the curves in figure~\ref{fig:decay_vs_ratio}, since $s = 1$ is the same for all values of $\hat{\omega}_\text{ini}$ for a fixed $\hat{\lambda}_\text{ini}$.

To study the effects of self-interactions for objects with the same core mass, we must use a different implementation of the scaling relations, as we shall discuss in the next section.

\subsection{Objects with known core mass and orbital period}\label{sec:fixed_Mc}

Before proceeding, it is important to note that one can use various definitions of the mass of the satellite once we have a density profile. For instance, one can use tidal mass $\hat{M}_\text{tidal}$ which is defined as the mass enclosed within the tidal radius $\hat{r}_\text{tidal}$. 
In our analysis, we use core mass, i.e. the mass contained within the radius at which density becomes half its central value (as defined in section~\ref{sec:decay_rates}) which can be uniquely found for every choice of $\hat{\lambda}_\text{ini}$ and $\hat{\omega}_\text{ini}$. 

\subsubsection{The procedure}

In order to study the effects of only self-interactions, we consider an object with a fixed $M_\text{fin}$ and $\omega_{\text{fin}}$. Then, for every $\hat{\lambda}_{\text{ini}}$, we require a scale value $s$ such that $\hat{M}_{\text{fin}} = s^{-1}\hat{M}_{\text{ini}}$ and $\hat{\omega}_\text{fin} = s^{-2}\hat{\omega}_{\text{ini}}$ both are satisfied. The procedure for this is as follows:

\begin{enumerate}
    \item Fix $\hat{M}_{\text{fin}}$ and $\hat{\omega}_{\text{fin}}$ from observations. 
    
    \item Fix the value of $\hat{\lambda}_{\text{ini}}$. For a chosen value of $\hat{\omega}_{\text{ini}}$ solve the dimensionless form of eqs.~(\ref{eq:mod_GP}) and~(\ref{eq:mod_poisson}). 

    \item Use $\hat{M}_{\text{fin}} = s^{-1}\hat{M}_{\text{ini}}$ to get a value of $s$, and check if the condition $\hat{\omega}_{\text{fin}} = s^{-2}\hat{\omega}_{\text{ini}}$ is satisfied within some desired accuracy ($5\%$ for our analysis). 

    \item If $\hat{\omega}_\text{fin} \neq s^{-2}\hat{\omega}_\text{ini}$, repeat steps $2$ and $3$ for a different value of $\hat{\omega}_{\text{ini}}$. Once the condition is satisfied for some $\hat{\omega}_\text{ini}$, using the corresponding numerical solution obtain $\hat{\Gamma}_{\text{ini}} = 2|\hat{\gamma}_I|$, and consequently obtain the scaled lifetime as $\tau = s^2\frac{\hbar}{mc^2}\hat{\Gamma}_\text{ini}^{-1}$. 
\end{enumerate}

A satellite with observed core mass $M_{\text{fin}}$ and orbital frequency $\omega_{\text{fin}}$ can be described by ULDM only if $\tau > t_\text{uni}$ where $t_\text{uni}$ is the age of the universe. 

\subsubsection{The absence and presence of self-interactions}\label{sec:absence_and_presence}

As a demonstration we have carried out the above procedure for a satellite whose core mass and orbital period are $M_{\text{fin}} = 9.1\times 10^7\ \text{M}_\odot$ and $T_{\text{fin}} =2\pi/\omega_\text{fin} = 1.6\ \text{Gyr}$ respectively (similar to values for UMi dwarf spheroidal from section~\ref{sec:fixed_rhoC}).

\begin{figure}[ht]
    \centering
    \begin{subfigure}[b]{0.45\textwidth}
        \centering
        \includegraphics[width = \textwidth]{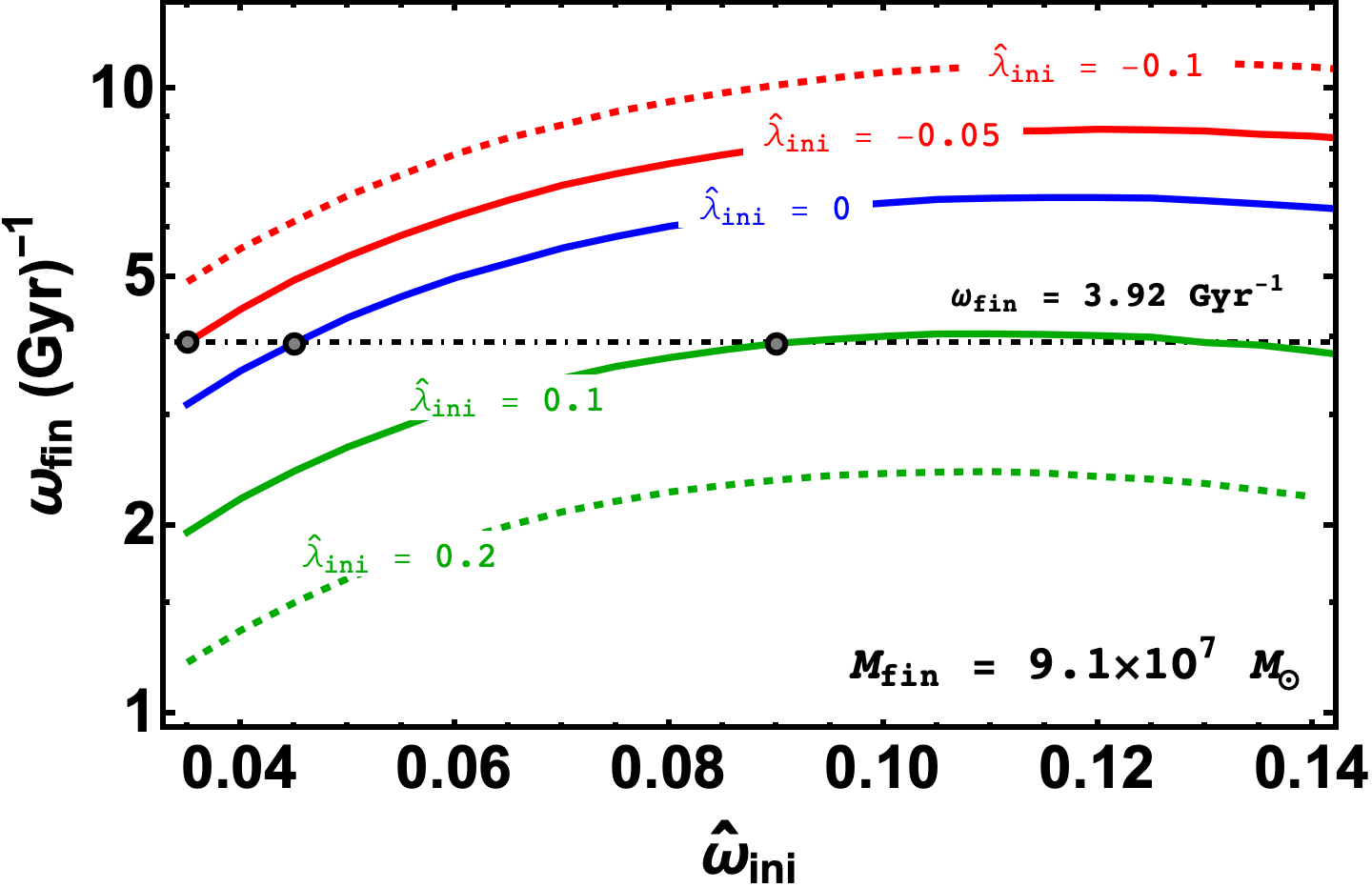}
        \caption{Possible values of $\omega_\text{fin}$ for a fixed $M_\text{fin}$ are shown. Each curve corresponds to a different $\hat{\lambda}_\text{ini}$. The horizontal gray line represents a fixed $\omega_\text{fin} = 3.92\ \text{Gyr}^{-1}.$ }
        \label{fig:fixed_core_mass_omega}
    \end{subfigure}
    \hfill
    \begin{subfigure}[b]{0.45\textwidth}
        \centering
        \includegraphics[width = \textwidth]{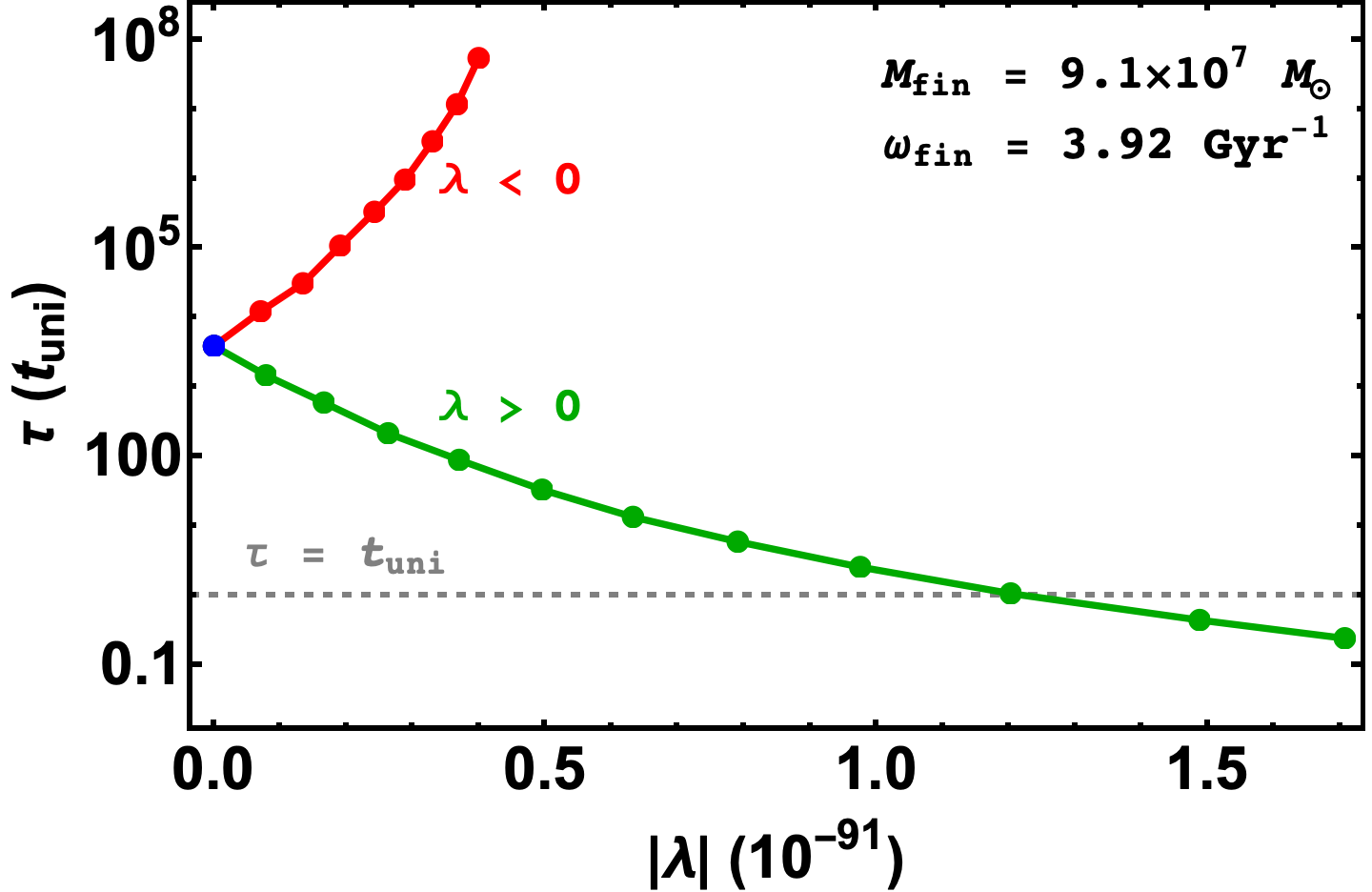}
        \caption{Lifetime in the units of the age of the universe is plotted for both attractive (red) and repulsive (green) self-interactions. The blue dot corresponds to the lifetime when $\lambda = 0$.}
        \label{fig:lifetime_fixed_core_mass}
    \end{subfigure}
    \caption{Left panel illustrates how one can model an object with a fixed core mass $M_\text{fin} = 9.1\times 10^7\ M_\odot$ and $\hat{\omega}_\text{fin} = 3.92\ \text{Gyr}^{-1}$. We plot how the lifetime changes in the presence of self-interactions for the same object in the right panel.}
    \label{fig:fixed_core_mass}
\end{figure}

It is important to note that not every combination of chosen values of $M_\text{fin}$ and $\omega_\text{fin}$ is allowed if one requires that both the scaling relations $M_\text{fin} = s^{-1}M_\text{ini}$ and $\omega_\text{fin} = s^{-2}\omega_\text{ini}$ are satisfied. As shown in figure~\ref{fig:fixed_core_mass_omega}, we find that for a fixed $\hat{\lambda}_{\text{ini}}$ and a chosen  $M_\text{fin}$, there is a maximum scaled $\omega_{\text{fin}}$ that can be obtained no matter which $\hat{\omega}_{\text{ini}}$ we start with. As a result, for the object with the above mentioned core mass and orbital frequency, self-coupling strengths with $\hat{\lambda}_{\text{ini}} > 0.1$ are not allowed, which is shown in figure~\ref{fig:fixed_core_mass_omega}. 

Following the procedure above, for the $\hat{\lambda}_\text{ini}$ values that can give rise to the observed values of $M_\text{fin}$ and $\omega_\text{fin}$, we have also plotted the lifetime of the satellite against $|\lambda|$ in figure~\ref{fig:lifetime_fixed_core_mass}. We find that for $\lambda = 0$, the lifetime of the satellite is $\tau \approx 3922\ t_\text{uni}$, which is already well above the minimum allowed lifetime $\tau = t_\text{uni}$. In the presence of attractive self-interactions, the lifetime increases as $|\lambda|$ increases shown by the red curve in figure~\ref{fig:lifetime_fixed_core_mass}. 
On the other hand, for ever larger repulsive self-interactions, lifetime of the satellite keeps on decreasing (shown by the green curve), until for $\lambda \gtrsim 1.2 \times 10^{-91}$,  $\tau < t_\text{uni}$ i.e., the satellite no longer survives until present day. Hence, if an object with $M_{\text{fin}} = 9.1\times 10^7 \ \text{M}_{\odot}$ and $\omega = 3.92\ \text{Gyr}^{-1}$ is observed, it will rule out $\lambda \gtrsim 1.2\times 10^{-91}$, imposing an upper limit on the strength of the self-coupling $\lambda$.

\subsubsection{Intuition for the effects of self-interactions}\label{sec:intuition}

As we have discussed in the previous section, the effect of self-interactions can be understood by looking at solutions for an object with a fixed core mass. Note that for stationary state solutions, we know that if soliton mass is fixed, central density is larger (as the soliton itself is squeezed) when self-interactions are attractive, while it is smaller (as the soliton expands) for repulsive self-interactions (see figure~1 in \cite{Chakrabarti:2022owq}). This is similar to the situation we have for the tunnelling problem, albeit for a leaky soliton. For instance, the object considered in figure~\ref{fig:fixed_core_mass}, has the orbital frequency $\omega_\text{fin} = 3.92\ \text{Gyr}^{-1}$ and core mass $M_\text{fin} = 9.1\times 10^7\ \text{M}_\odot$. The scaled central density for this object when $\lambda = 0$ turns out to be $\rho_c \approx 0.13\ \text{M}_\odot/\text{pc}^{3}$. On other hand, for $\lambda = -2.88\times 10^{-92}$, the central density is $\rho_c = 0.22\ \text{M}_\odot/\text{pc}^{3}$ while for $\lambda = 1.48\times 10^{-91}$, $\rho_c \approx 0.03\  \text{M}_\odot/\text{pc}^{3}$. From the discussion in section~\ref{sec:comparison} we also know that a larger (smaller) central density implies a larger (smaller) lifetime, which explains the effect of attractive and repulsive self-interactions.

Further, we have plotted potential barriers for three values of $\hat{\lambda}_\text{ini} = \{-0.05, 0, 0.1\}$ for the above mentioned object in figure~\ref{fig:SI_barrier} to confirm our intuition. To satisfy scaled quantities, the value of $\hat{\omega}_\text{ini}$ required is smaller (larger) for attractive (repulsive) self-interactions, leading to a wider (narrower) potential barrier. Hence, in the presence of attractive self-interactions, the satellite survives for longer while for repulsive self-interactions, the opposite is true.

We find that the approximate approach in appendix~\ref{app:approx_appraoch} holds up qualitatively for satellites with constant core masses as well. As seen in figure~\ref{fig:wkb_fixed_core_mass}, the approximate potential barriers exhibit the same behaviour as the numerically solved barriers in figure~\ref{fig:SI_barrier}. This is because even in the absence of tidal effects a $\hat{\lambda}_{\text{ini}} < 0$ ($\hat{\lambda}_{\text{ini}} > 0$) will lead to a smaller (larger) initial core mass. Hence, one requires a different $\hat{\omega}_{\text{ini}}$ for different $\hat{\lambda}_\text{ini}$ to obtain the same core mass. 
However, it is worth noting that the numbers obtained from the approximate approach may not be accurate since the system is not numerically solved in the presence of the tidal potential (see appendix~\ref{app:approx_appraoch} for more details).

\subsection{Saving ULDM?}\label{sec:saving_ULDM}

From the discussion in the previous sections it is clear that repulsive self-interactions make the satellite even more susceptible to tidal disruption, thereby decreasing lifetime. In this section, we shall discuss how attractive self-interactions can help the case of ULDM for a fixed $m$, by looking at two different objects. 

\begin{figure}[ht]
    \centering
    \begin{subfigure}[b]{0.45\textwidth}
        \centering
        \includegraphics[width = \textwidth]{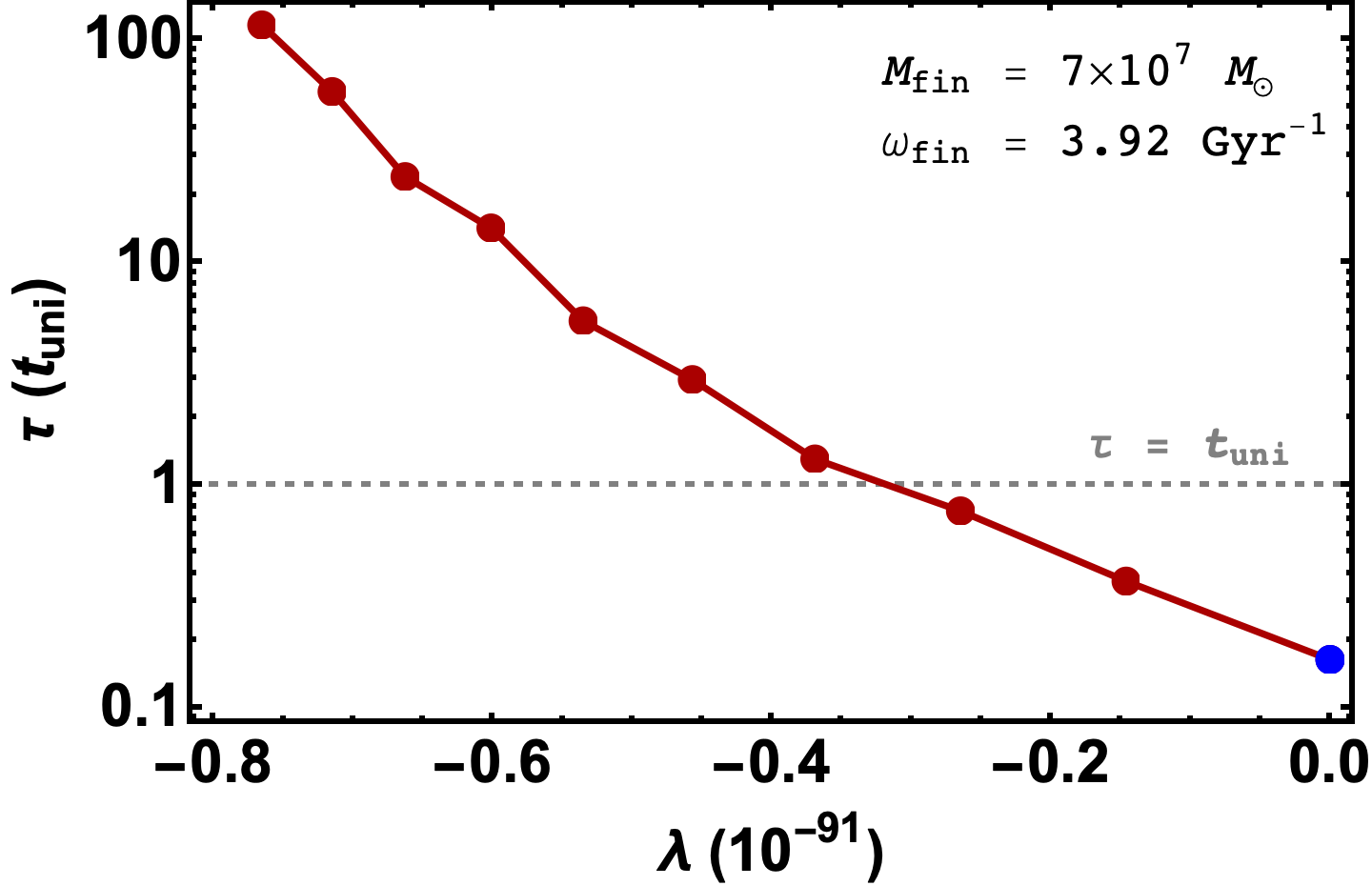}
        \caption{If this hypothetical object is to be described within the ULDM paradigm with $m =10^{-22}\ \text{eV}$, attractive self-interactions with $\lambda \lesssim -0.32\times 10^{-91}$ are required for it survive until present day.This can be used to impose constraints on self-interactions for a fixed $m$, as we see in the right panel}
        \label{fig:fixed_Mc_att_lambda}
    \end{subfigure}
    \hfill
    \begin{subfigure}[b]{0.45\textwidth}
        \centering
        \includegraphics[width = \textwidth]{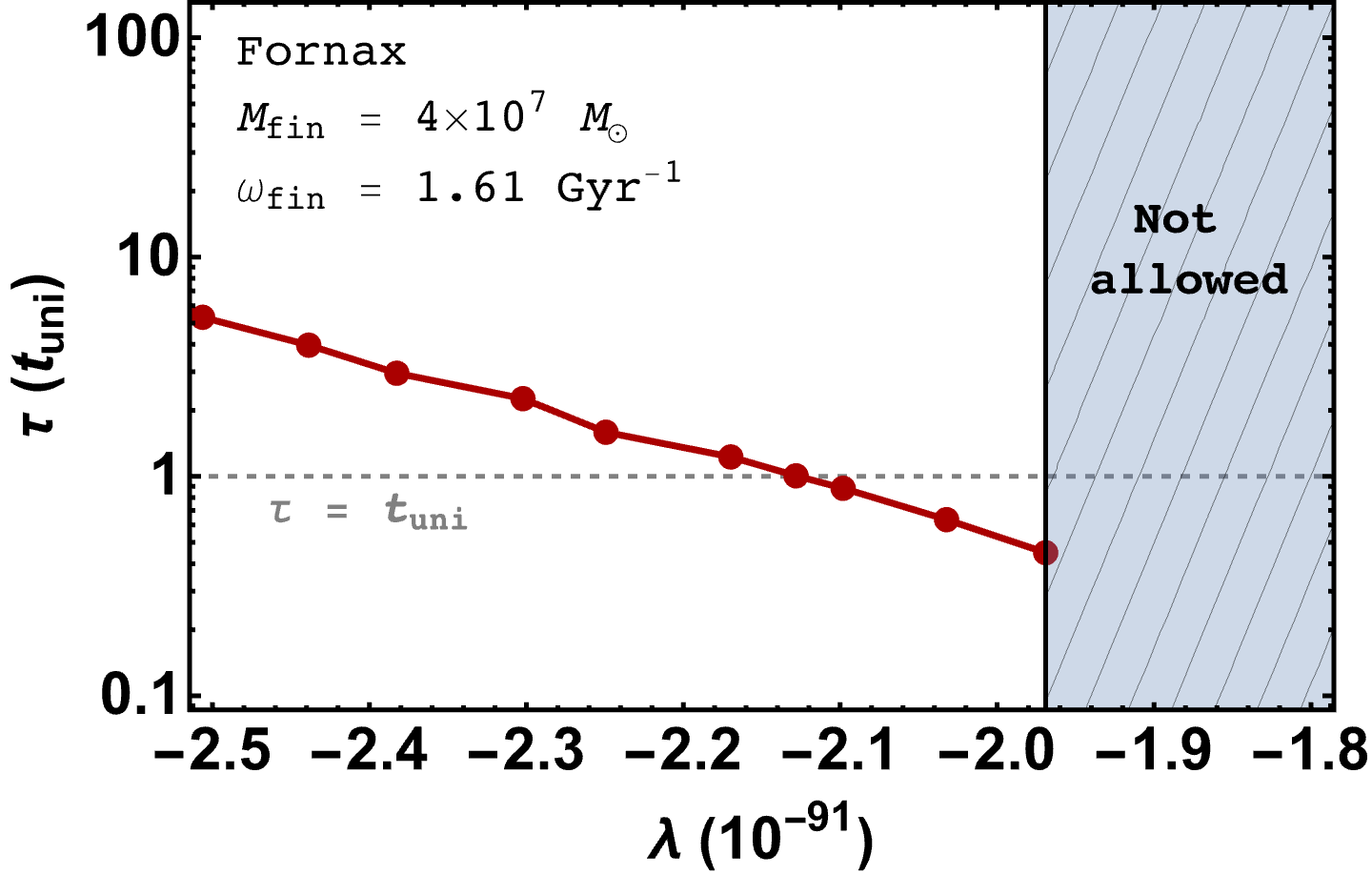}
        \caption{ULDM with $m = 10^{-22}\ \text{eV}$ and $\lambda > -1.9\times 10^{-91}$ cannot model the Fornax dwarf galaxy. Further, for $-2.12\times 10^{-91} < \lambda < -1.97\times 10^{-91}$ while we can model it, it will not survive until present day. For $\lambda < -2.12\times 10^{-91}$ however, its lifetime will be larger than the age of the universe.}
        \label{fig:saving_fornax}
    \end{subfigure}
    \caption{Lifetime of the satellite galaxy versus strength of attractive self-interactions are plotted for a hypothetical object (left panel) and Fornax dwarf galaxy (right panel).}
    \label{fig:saving_ULDM}
\end{figure}

Firstly, let us consider an object with orbital period $T_{\text{fin}} = 1.6\ \text{Gyr}$ and a core mass $M_\text{{fin}} = 7\times 10^7\ \text{M}_\odot$. For this object, if we follow the procedure from the previous section, we find that in the FDM regime (i.e. $\lambda = 0$) the  lifetime of the satellite is smaller than the age of universe: $\tau \approx 0.16\ t_\text{uni}$. Here, repulsive self-interactions are only going to make things worse, by reducing the lifetime even further. However, from figure~\ref{fig:lifetime_fixed_core_mass}, it is clear that attractive self-interactions should help and we find that they indeed do, as shown in figure~\ref{fig:fixed_Mc_att_lambda}. 

As $\lambda$ becomes more negative, the lifetime of the satellite increases until $\lambda \lesssim -0.32\times 10^{-91}$, when $\tau \gtrsim t_\text{uni}$ and the object can survive until present day. Hence, if an object with core mass $\sim 7\times 10^7\ \text{M}_\odot$ and orbital frequency $\sim 3.92\ \text{Gyr}^{-1}$ is observed, while $\lambda = 0$ will be ruled out, $\lambda < 0$ can still account for the survival of the satellite. 

We then look at another object, viz. the Fornax dwarf spheroidal, whose core mass and orbital period are $M_\text{fin} = 4\times 10^7\ M_\odot$ and $T_\text{fin} = 2\pi/\omega_\text{fin} = 3.9\ \text{Gyr}$ \cite{Hertzberg:2022vhk}. For $m = 10^{-22}\ \text{eV}$ the $\hat{\lambda}_\text{ini} = 0$ case cannot model Fornax (following a similar argument to the case for $\hat{\lambda}_\text{ini} > 0.1$ for the object in figure~\ref{fig:fixed_core_mass_omega}).
This should be compared with results in \cite{Hertzberg:2022vhk} where it is claimed that the existence of Fornax dwarf galaxy implies that $m = 10^{-22}\ \text{eV}$ is ruled out for FDM. 

From figure~\ref{fig:saving_fornax}, one can see that Fornax can be described only if the self-interaction strength $\lambda \approx -1.97\times 10^{-91}$. However, the corresponding scaled lifetime of Fornax will be smaller than the age of the universe. As we consider increasingly negative $\lambda$ we again see that the lifetime increases until, for $\lambda \lesssim -2.12\times 10^{-91}$, the lifetime becomes larger than the age of the universe, allowing Fornax to survive until present day. This implies that if ULDM mass is to be $m = 10^{-22}\ \text{eV}$ then the attractive self-interactions $\lambda \lesssim -2.12\times 10^{-91}$ are necessary to explain the survival of Fornax until present day. 

Also note the slight wobbly behaviour of the curves in figure~\ref{fig:saving_ULDM}. Our understanding is that this could be because: (a) for all $\hat{\lambda}_{\text{ini}}$ and $\hat{\omega}_\text{ini}$ values, we have fixed $\hat{r}_{m}$ (the distance at which we match the numerical solution with the WKB one - see discussion around eq.~(\ref{eq:BCs})) to be $55$, (b) the numerical solution becomes highly oscillatory at large $\hat{r}$ (see eq.~(\ref{eq:WKB})). We also found that this wobble appears to occur at larger values of $\hat{\lambda}_{ini}$ (which require smaller values of $\hat{\omega}_{ini}$ to describe objects with the same scaled core mass).

Hence, we have demonstrated that one can impose constraints on the strength of attractive self-interactions of ULDM from observations of DM core masses and orbital periods of satellite galaxies.

\subsection{Comparison with Ref~\cite{Glennon:2022huu}}
Recently, the effects of self-interactions have been studied by carrying out full time-dependent simulations in \cite{Glennon:2022huu}. Here, authors simulate solitons with total mass $M_s$ in the narrow range $1.04 \times  10^8\ M_\odot - 1.13\times 10^8\ M_\odot$ orbiting at the outskirts of a halo ($M_h\sim 10^{11}\ M_\odot$) in the presence of attractive and repulsive self-interactions with $\lambda \approx \pm 2\times 10^{-92}$. 
Note that here the authors deal with a full time-dependent problem, and simulate the evolution of the solitons for multiple orbits. 
They find that going from $\lambda = 0$ to $\lambda = \pm 2.4\times 10^{-92}$ can increase or decrease disruption time by up to a factor of $\sim 1.5$. 

Using quasi-stationary solutions, we find that for core mass $M_{\text{fin}} = 7\times 10^7\ M_\odot$ (which is not the total soliton mass) and $\omega_\text{fin} = 3.92\ \text{Gyr}^{-1}$, lifetime of the satellite galaxy will be larger by a factor of $\sim 4$ for $\lambda = -2.4\times 10^{-92}$ compared to $\lambda = 0$ (see figure~\ref{fig:fixed_Mc_att_lambda}).

Further, for a larger core mass $M_\text{fin} = 9.1\times 10^7\ M_\odot$, we find that the effects of SI are larger, and the lifetime of the satellite increases by two orders of magnitude as one goes from $\lambda = 0$ to $\lambda = -2.4\times 10^{-92}$.
On the other hand, going from no self-interactions to $\lambda = 2.4\times 10^{-92}$ decreases the lifetime by an order of magnitude (see figure~\ref{fig:lifetime_fixed_core_mass} for curves considering both attractive and repulsive self-interactions). Hence, our results agree with those of \cite{Glennon:2022huu} : attractive self-interactions lead to longer-lived satellite galaxies compared to the absence of self-interactions, while repulsive self-interactions lead to shorter-lived satellites. However, the extent of the effect of $\lambda\neq 0$ differs. 

It should be noted that apart from working with quasi-stationary solutions there are some key differences in our treatment of the problem compared to Ref.~\cite{Glennon:2022huu}:
\begin{itemize}
    \item In our work, we cannot reliably define a total soliton mass since the integral in eq.~(\ref{eq:soliton_mass}) does not converge at large $r$. Hence, it is not clear if we are looking at similar soliton masses to \cite{Glennon:2022huu}. The orbital periods (assuming a circular orbit) that the authors consider are also greater than what we can probe for the core masses that we consider. This is because in our method, the scaling relations in eq.~(\ref{eq:scaling}) restrict the combinations of $M_\text{fin}$ and $\omega_\text{fin}$ that can be probed. 
    \item In \cite{Glennon:2022huu}, the authors initialize the solitons with a fixed $\rho_c/\rho_H \approx 50$ for all $\lambda$ values. However, in our work, we utilize scaling relations in eq.~(\ref{eq:scaling}) to describe realistic solutions which requires different $\hat{\omega}_\text{ini}$ values to probe a fixed scaled core mass and orbital period for different $\lambda$ values. This will lead to a different ratio $\rho_c/\rho_H$ for every $\lambda$ value which can explain the larger effect of self-interactions in our case.  
\end{itemize}

\section{Discussion and Conclusion}\label{sec:summary}

Dark Matter (DM) could be composed of elementary or composite particles of a single species. Since there are no elementary or composite particles in the standard model of particle physics with properties expected from DM, it is important to look for theories beyond the standard model. Thus, astrophysical and cosmological observations of DM could teach us something about new physics.

If the particles constituting DM are Bosons, for sufficiently small particle masses, the number density is allowed to be so large that a classical wave description is valid (see \cite{Allali:2020ttz, Allali:2020shm, Allali:2021puy, Herdeiro:2022gzp, Eberhardt:2023axk, Chakrabarty:2021yzg, Eberhardt:2021okc, Eberhardt:2022rcp, Alcubierre:2022rgp, Proukakis:2023nmm, Proukakis:2023kcd} and references therein).
This so-called Ultra-Light Dark Matter (ULDM) or Fuzzy Dark Matter (FDM) (in the absence of self-interactions) will lead to unique wave-like observational signatures at galactic scales while behaving like Cold Dark Matter (CDM) at larger scales \cite{Matos:1999et, Lee:1995af, Hu:2000ke, Schive:2014dra} (also see \cite{Ferreira:2020fam, Hui:2021tkt, 2168507} for recent reviews), and can potentially solve small-scale issues with CDM \cite{Bullock:2017xww}. However in recent years, much of the relevant mass range for ULDM with no self-interactions $\left[10^{-24}\ \text{eV}, 10^{-20}\ \text{eV}\right]$ has been severely constrained using various astrophysical and cosmological observations. See for instance constraints from rotation curves \cite{Bar:2021kti, Khelashvili:2022ffq, Banares-Hernandez:2023axy}, Lyman-$\alpha$ data \cite{Irsic:2017yje, Kobayashi:2017jcf, Rogers:2020ltq}, CMB and Large Scale Structure \cite{Hlozek:2017zzf, Lague:2021frh}, dynamical heating of stars in ultra-faint dwarf galaxies \cite{Dalal:2022rmp}, as well as recent results from NANOGrav and Pulsar Timing Arrays \cite{NANOGrav:2023hvm, Smarra:2023ljf}. 

There is however a possible out, in the form of non-negligible self-interactions. For ULDM, many of the constraints mentioned earlier are obtained in the FDM regime, i.e. the effect of self-interactions is taken to be negligible. As we have mentioned earlier, presence of extremely small self-interactions can drastically impact solutions of the Gross-Pitaevskii-Poisson system, which can potentially alter the above-mentioned constraints. For instance, the stringent constraints imposed by Lyman-$\alpha$ data can potentially be relaxed if the effects of self-interactions are taken into account (see discussion below eq.~(3.45) in section~3.2.3 in \cite{2168507}, section~5.1.2 in \cite{Ferreira:2020fam} as well as \cite{Zhang:2017dpp, Leong:2018opi}). Indeed, there has been a growing interest in recent years to study and constrain self-interactions of scalar field dark matter \cite{Li:2013nal, Dev:2016hxv, Suarez:2016eez, Desjacques:2017fmf, Cembranos:2018ulm, Delgado:2022vnt, Chakrabarti:2022owq, Mocz:2023adf, Dave:2023wjq, Ravanal:2023ytp, Kadota:2023wlm, Budker:2023sex, Boudon:2023qbu, Winch:2023qzl}. 

In this paper, we studied the effects of self-interactions for a system consisting a satellite dwarf galaxy orbiting the centre of a host halo at a distance $a$ with an orbital period $T = 2\pi/\omega$ where $\omega$ is the orbital frequency of the satellite. As shown by \cite{Hertzberg:2022vhk}, the tidal effects due to the halo will lead to an external potential of the form $-\frac{3}{2}m\omega^2r^2$ in the time-independent Schrödinger-Poisson system. Note that the authors consider only the $\ell = 0$ mode since higher order modes contribute very little for small values of $\omega$. It is also important to note that the factor of $\frac{3}{2}$ is valid only if the satellite is orbiting far enough away from the halo centre. Assuming that one can introduce a similar term to the Gross-Pitaevskii-Poisson equations (i.e. equations of motion for a classical scalar field with self-interactions), we obtain a time-independent system given by eqs.~(\ref{eq:mod_GP}) and~(\ref{eq:mod_poisson}). 

One can solve this system by treating it as a quantum mechanical tunnelling problem as authors in \cite{Hui:2016ltb, Du:2018qor, Hertzberg:2022vhk} have done albeit for $\lambda = 0$. To model tunnelling in the time-independent approximation, one requires quasi-stationary states which arise as a consequence of outgoing wave boundary conditions describing a scattering-like state (given by eq.~(\ref{eq:WKB}) at a large distance from the origin viz. the centre of the satellite in our case). Note that this forces the eigenvalue $\gamma$ to be complex where $\text{Im}(\gamma) \equiv \gamma_I$ characterises the rate of tunnelling while its inverse gives us the timescale over which the satellite loses its mass through tunnelling. To solve the system, we looked for the correct values of $\gamma_R$ and $\gamma_I$ such that the solution matches with the outgoing wave solution at large $r$ (see section~\ref{sec:numerical_procedure} for details). 

An important point worth noting here is that our approach attempts to solve a time-dependent problem using a time-independent description, in the sense that mass of DM in the satellite decays exponentially, i.e. $M(t) = M_0e^{-\Gamma t/\hbar}$ where $\Gamma$ is constant. However, detailed modelling \cite{Du:2018qor} based on explicit solution of time-dependent SP system indicates that the time-dependence of the mass of DM in the satellite could be different from exponential. It is worth noting that the effects of self-interactions have been considered in recent simulations \cite{Glennon:2022huu} where the authors tackle a full time-dependent problem. The results we have arrived at by working with the exponential model agree well with the results of their simulations, in that attractive (repulsive) self-interactions lead to a longer (shorter) lifetime for a satellite dwarf orbiting the host halo. 

We have then solved the modified time-independent GPP system numerically to obtain decay rates for various values of dimensionless $\hat{\lambda}_\text{ini}$ and $\hat{\omega}_\text{ini}$ for a fixed $m = 10^{-22}\ \text{eV}$. Using scaling relations in eq.~(\ref{eq:scaling}) to describe satellite galaxies with typical core masses and orbital periods, we found that for a satellite with core mass $M_c \sim 10^8\ M_\odot$, and orbital period $T\sim \mathcal{O}(\text{Gyr})$, self-couplings as small as $\lambda \gtrsim \pm 10^{-92}$ can have a substantial effect on its lifetime (see section~\ref{sec:SI_impact} for details). We looked at an object with a given core mass and orbital period and found that as we increase the strength of repulsive self-interactions, the lifetime of the satellite becomes shorter while the opposite is true for attractive self-interactions (see figure~\ref{fig:fixed_core_mass}).

Thus, one can place upper limits on repulsive self-interactions by requiring that the satellite survive at least as long as the age of the universe (see section~\ref{sec:fixed_Mc}). This is because repulsive self-interactions are responsible for a lower central density and a narrower potential barrier leading to short-lived satellite compared to no self-interactions. On other hand, we found that attractive self-interactions which lead to higher central densities and wider potential barriers, allow for a satellite to survive for longer compared to the case of no self-interactions (see figure~\ref{fig:SI_barrier} and the discussion in section~\ref{sec:intuition}). Assuming no DM self-interactions, if an object of known core mass and orbital period has a lifetime much less than the age of the universe, then introduction of sufficiently strong attractive self-interactions will make its lifetime large enough to survive until present day (see figure~\ref{fig:fixed_Mc_att_lambda}). This can severely modify the conclusions of \cite{Hertzberg:2022vhk} which assumes negligible self-interactions. 

In \cite{Dave:2023wjq}, we had found that (a) repulsive self-interactions with $\lambda \gtrsim 10^{-91}$ can explain observed rotation curves for dark matter dominated galaxies as well as satisfy an appropriate soliton-halo relation, and, (b) including attractive self-interactions did not have the same interesting effect. On the contrary, in the present work, we found that consideration of lifetimes of satellite galaxies prefers attractive self-couplings of comparable strengths. 

Ultra light scalars like the ones considered in this paper could turn up in many ultraviolet completions of the standard model (see e.g. \cite{Cicoli:2021gss, Hamaide:2022rwi, Cicoli:2023opf, Gendler:2023kjt} for some recent papers and references therein). If observations can be used to constrain the self-couplings of such scalars, this can go a long way towards uncovering physics which operates at ultra-short distances. It is well-known that axions will have a negative quartic self-coupling - in particular, ultra-light axions can have extremely small self-couplings. Thus the results of this paper are intriguing, given recent interest in axion dark matter \cite{AlfredAmruth:2023mel}.

\acknowledgments
This work is supported by Department of Science and Technology, Government of India under Indo-Russian call for Joint Proposals (DST/INT/RUS/RSF/P-21). 
BD acknowledges support from the above mentioned project as a Junior Research Fellow.
This work is also supported by Department of Science and Technology, Government of India under Science and Engineering Research Board - State University Research Excellence (SUR/2022/005391).
BD would like to thank Arka Banerjee (IISER Pune) for discussions at the initial stage of the work. 
This research was also supported in part by the International Centre for Theoretical Sciences (ICTS) for participating in the program - Less Travelled Path to the Dark Universe (code: ICTS/ltpdu2023/3).

\appendix

\section{Approximate Approach}\label{app:approx_appraoch}

In the main body of this paper, in order to solve GPP equations modified by the presence of tidal potential, we took the approach of quasi-stationary states involving complex energy and outgoing wave boundary conditions.
In standard textbooks, one often deals with tunneling problems (to study e.g. $\alpha$-decay of nuclei) by using WKB approximation which involves real energy (see e.g. \cite{Griffiths:2018qm}).
It is a good idea to attempt to analyse the current problem in the usual textbook approach, this is done in this section.
Before proceeding, we must state the basic assumption for this method: We do not solve the system numerically in the presence of the tidal-potential. Instead, we solve with GPP system with $\hat{\omega}_{\text{ini}} = 0$ to obtain the stationary-state eigenvalue $\hat{\gamma}$ (corresponding to a node-less solution), $\hat{\phi}(\hat{r})$ and $\hat{\Phi}_{SG}(\hat{r})$. Now we construct an approximate effective potential 
    \begin{equation}\label{eq:eff_pot}
        \hat{V}_{\text{eff}} = \hat{\Phi}_{SG} + \hat{\Phi}_H + \hat{\Phi}_{SI}\ ,
    \end{equation}
where $\hat{\Phi}_{SI} \equiv 2\hat{\lambda}_{\text{ini}}|\hat{\phi}(\hat{r})|^2$ and $\hat{\Phi}_H \equiv -\frac{3}{2}\hat{\omega}_{\text{ini}}^2\hat{r}^2$ with some $\hat{\omega}_{\text{ini}} > 0$. It is important to note that in reality the presence of a tidal term in the Scrh\"{o}dinger equation should alter the density profile $|\hat{\phi}(\hat{r})|^2$ as well as the ground state energy $\hat{\gamma}$ (see figure~\ref{fig:example_sol}). This in-turn will alter $\hat{\Phi}_{SG}$. However, for small values of $\hat{\omega}$, $\hat{V}_{\text{eff}}$ as we have defined it, should serve as a good enough approximation. 
    
To understand the effect of $\hat{\omega}_{\text{ini}}$, let us construct $\hat{V}_\text{eff}$ for $\hat{\lambda}_{\text{ini}} = 0$ and different values of $\hat{\omega}_{\text{ini}}$. The resultant potential barriers are shown in figure~\ref{fig:approx_pot_barrier}. It is clear that a larger $\hat{\omega}_{\text{ini}}$ will lead to a narrower barrier, making it easier to tunnel through, while a smaller $\hat{\omega}_{\text{ini}}$ will widen the barrier until $\hat{\omega}_{\text{ini}}= 0$ for which there will no tunnelling. Figure~\ref{fig:changing_barriers} also illustrates that a small enough $\hat{\omega}_{\text{ini}}$ will have a negligible effect on the total potential in the inner region. Treating the soliton as a particle with energy $\hat{\gamma}$, for every $\hat{\omega}_{\text{ini}}$, the width of the potential barrier experienced by the particle will be $\hat{r}_2 - \hat{r}_1$ where $\hat{\gamma} = \hat{V}_{\text{eff}}$ at $\hat{r}_1$ and $\hat{r}_2$. The transmission probability $\hat{T}_p$ at $\hat{r}_2$ following \cite{Griffiths:2018qm} is then simply 

\begin{equation}\label{eq:transmission}
    \hat{T}_p = e^{-2\hat{\sigma}}\ ,
\end{equation}
where $\hat{\sigma}$ is the integral

\begin{equation}
\hat{\sigma} = \int_{\hat{r}_1}^{\hat{r}_2}\sqrt{2\left(\hat{V}_{\text{eff}}(\hat{r}) - \hat{\gamma}\right)}d\hat{r}\ .
\end{equation}
It is worth noting that $\hat{\sigma}$ is invariant under scaling (this can be seen by applying scaling relations from eq.~(\ref{eq:scaling})). Now, assuming that the particle is moving around in the potential well hitting the barrier at $\hat{r}_1$ repeatedly, the frequency of collisions can be written as $\hat{v}/2\hat{r}_1$, where $\hat{v}$ is the velocity of the particle. A product of the transmission probability and the collision frequency is the probability of tunnelling per unit time i.e. the decay rate $\hat{\Gamma}. $\footnote{Note that $\hat{\Gamma} = \frac{\hbar}{mc^2}\Gamma$.} Since $\hat{\Gamma}$ is the inverse of time, it scales as $\Gamma \rightarrow s^{-2}\Gamma$ (from eq.~(\ref{eq:scaling})). Since any change in the exponent will dominate over any change in the velocity, one can approximate the decay rate as 

\begin{equation}\label{eq:wkb_decay}
    \hat{\Gamma} = s^{-2}\frac{\hat{v}}{2\hat{r}_1}e^{-2\hat{\sigma}} \approx s^{-2}e^{-2\hat{\sigma}}\ .
\end{equation}
Here we have ignored the $\frac{\hat{v}}{2\hat{r}_1}$ factor since we do not have an estimate for the velocity of the 'particle' in this case. Moreover, the exponential will have the dominant contribution to the decay rate.

\subsection{Negligible self-interactions} 

\begin{figure}[ht]
    \centering
    \includegraphics[width = 0.65\textwidth]{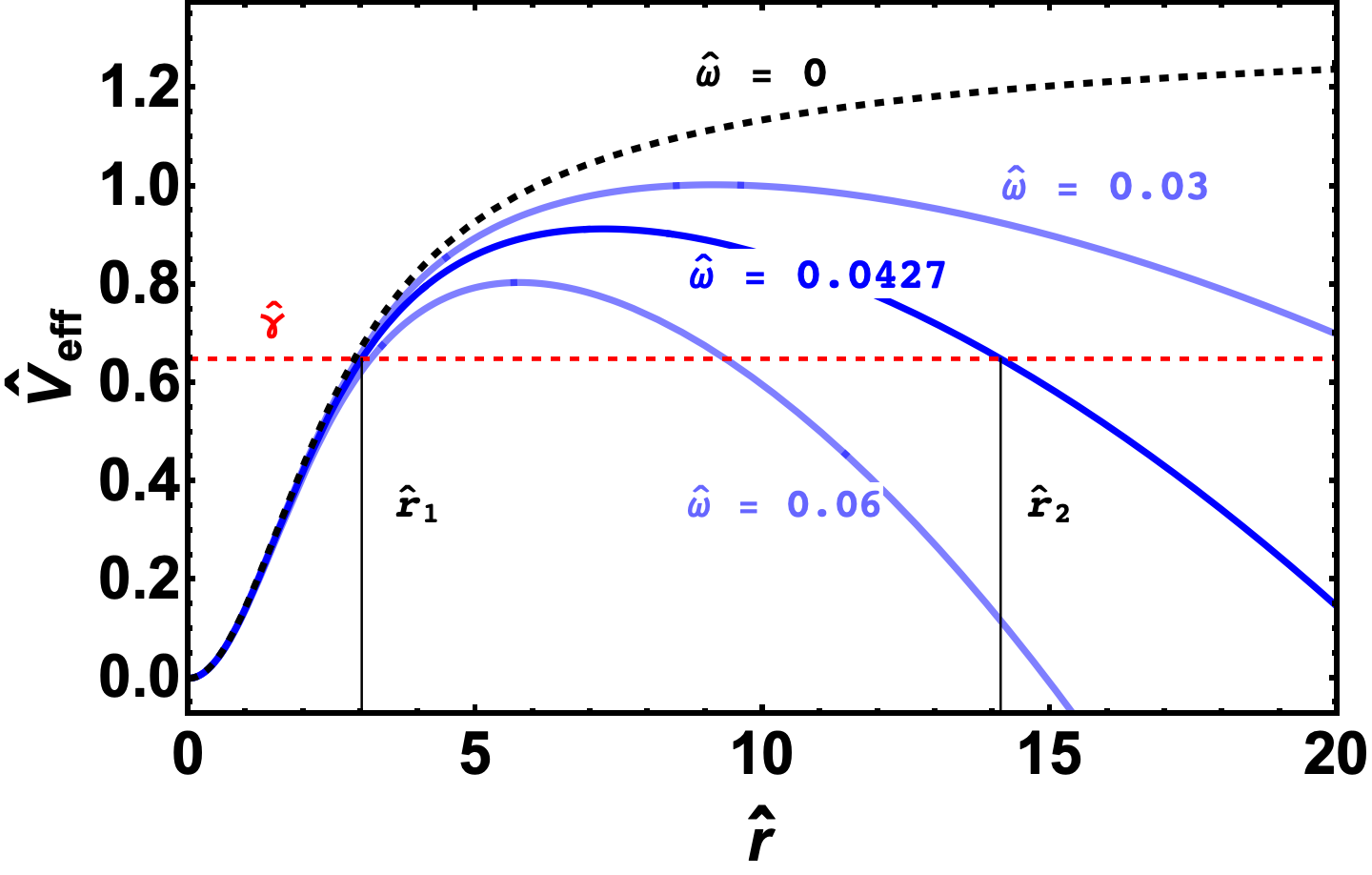}
    \caption{Approximate effective potentials for various values of $\hat{\omega}_\text{ini}$ are shown. For the UMi spheroidal, required value of $\hat{\omega}_{\text{ini}} = 0.0427$ is shown by the dark blue curve. $\hat{d} = \hat{r}_2 - \hat{r}_1$ is the width of the barrier while the red line is the ground state energy eigenvalue in the absence of the tidal potential.}
    \label{fig:approx_pot_barrier}
\end{figure}

Let us apply this approach to a real system, that of the `UMi' spheroidal, whose central density is $\rho_c \approx 0.15\ M_\odot \text{pc}^{-3}$ and the orbital period is $T \approx 1.6\ \text{Gyr}$, where $\omega = 2\pi/T$ \cite{Hertzberg:2022vhk}. Following the discussion in section~\ref{sec:fixed_rhoC}, the scale value $s = 7225$ and $\hat{\omega}_{\text{ini}} = 0.0427$ are obtained. The resultant approximate effective potential is shown in figure~\ref{fig:approx_pot_barrier}. The decay rate using eq.~(\ref{eq:wkb_decay}) (after scaling) is then estimated to be 

\begin{equation}
    \Gamma \sim 10^{-3}/\text{t}_{\text{uni}}\ ,
\end{equation}
which is an order of magnitude larger than the one obtained in \cite{Hertzberg:2022vhk}. As discussed there are several sources of uncertainty: (a) The self-gravitational potential $\hat{\Phi}_{SG}$ will be altered in the presence of the tidal potential, (b) The extra term in the Schr\"{o}dinger equation will also impact the eigenvalue $\hat{\gamma}$. If either of these two quantities change even a little, the width of the barrier experienced by the soliton might change which will lead to a drastically different decay rate because of the exponential dependence on $\hat{\sigma}$. This implies that this approximation can only work when $\hat{\omega}_{\text{ini}}$ is small enough to not appreciably affect the solution of the SP system. Further, we have also ignored $\hat{v}/2\hat{r}_1$ factor, which can lead to further alterations to the decay rate. 

However, as we have now seen, even a crude toy-approach using the WKB approximation is useful for qualitatively understanding the impact of $\hat{\omega}$. This is also true for the effects of self-interactions (which we shall discuss in section~\ref{sec:SI_toy}).

\subsection{Non-negligible self-interactions}\label{sec:SI_toy}

In this section, similar to what we did in the main paper we shall look at two cases: (a) fixed scaled core density and (b) fixed scaled core mass. For both cases we look at three values of $\hat{\lambda}_\text{ini} = \{-0.05, 0, 0.1\}$. We follow the same procedure for obtaining the appropriate scale value as we did in sections~\ref{sec:fixed_rhoC} and~\ref{sec:fixed_Mc} and find:

\begin{figure}[ht]
    \centering
    \begin{subfigure}[b]{0.45\textwidth}
        \centering
        \includegraphics[width = \textwidth]{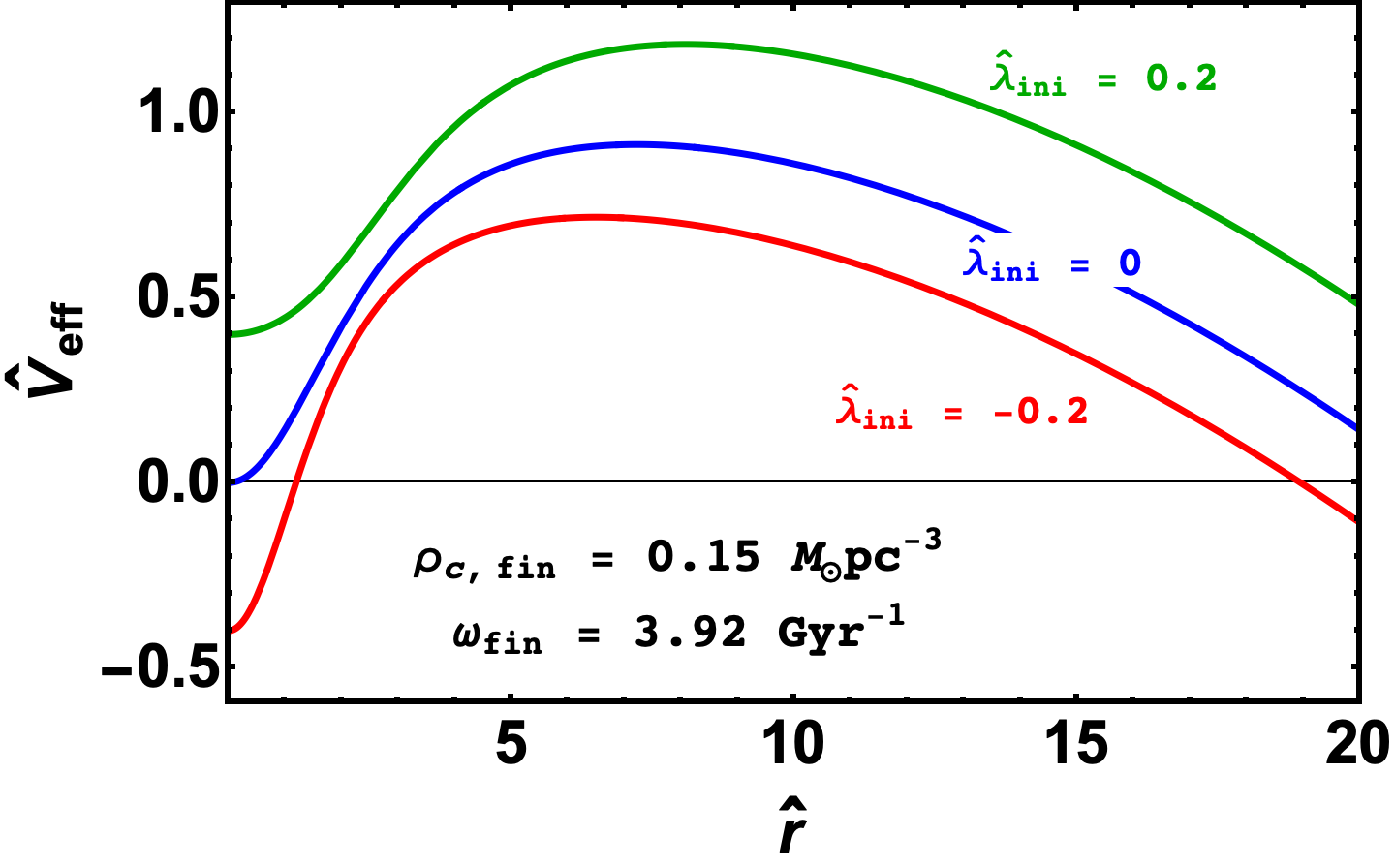}
        \caption{Approximate effective potentials for three values of $\hat{\lambda}_\text{ini}$ are shown, for a fixed $\hat{\omega}_\text{ini} = 0.0427$ and $s = 7225$ which corresponds to a scaled $\rho_{c,\text{fin}}= 0.15\ M_\odot\text{pc}^{-3}$ and $\omega_\text{fin} = 3.92\ \text{Gyr}^{-1}$.}
        \label{fig:wkb_fixed_rhoc}
    \end{subfigure}
    \hfill
    \begin{subfigure}[b]{0.45\textwidth}
        \centering
        \includegraphics[width = \textwidth]{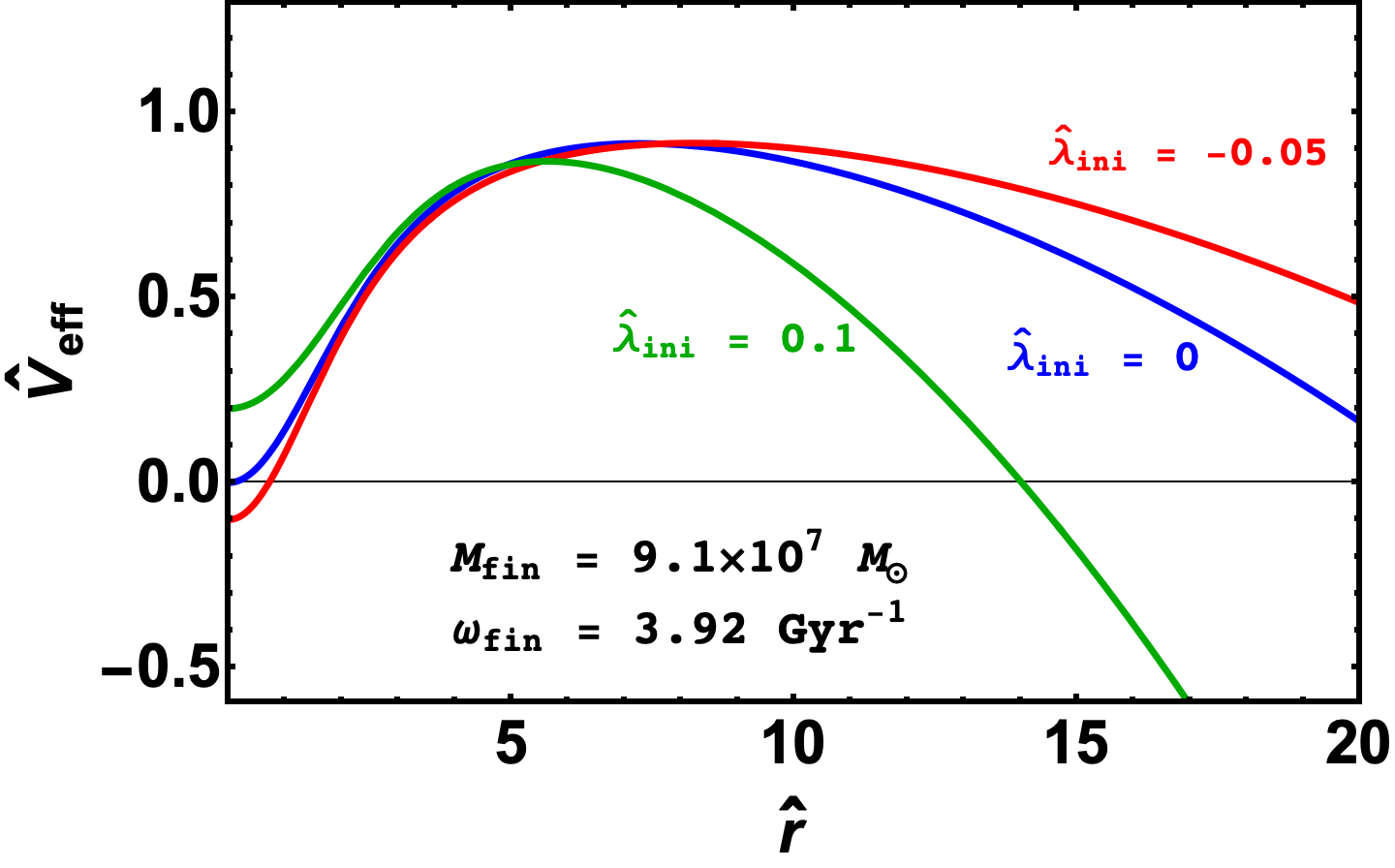}
        \caption{Approximate effective potential $\hat{V}_\text{eff}$ for different values of $\hat{\omega}_{\text{ini}}$ and $\hat{\lambda}_\text{ini}$ such that scaled $M_\text{fin} = 9.1\times 10^7\ M_\odot$ and $\omega_\text{fin} = 3.92\ \text{Gyr}^{-1}$. The barriers are similar to what we get from numerical solutions in figure~\ref{fig:SI_barrier}.}
        \label{fig:wkb_fixed_core_mass}
    \end{subfigure}
    \caption{Left panel illustrates the effects of self-interactions on the effective potential for objects with the same central density and orbital period. In the right panel, we consider different self-interaction strengths for objects with the same core and orbital period.}
    \label{fig:approx_SI_barriers}
\end{figure}

\begin{itemize}
    \item The scaled central density and orbital frequency are fixed to be $\rho_{c, \text{fin}} = 0.15\ M_\odot\text{pc}^{-3}$ and $\omega_\text{fin} = 3.92\ \text{Gyr}^{-1}$. The corresponding scale value is fixed for all $\hat{\lambda}_\text{ini}$ to be $s = 7225$ while $\hat{\omega}_\text{ini} = 0.0427$. For $\hat{\lambda}_\text{ini} = 0$ the scaled lifetime is $\tau \approx 235\ t_\text{uni}$. For $\hat{\lambda}_\text{ini} = \pm0.2$ the corresponding scaled $\lambda = \pm 1.41\times 10^{-91}$ while the scaled lifetimes are found to be $\tau \approx 6387\ t_\text{uni}$ and $\tau \approx 36.60\ t_\text{uni}$ for repulsive and attractive self-interactions respectively.  

    \item For $M_\text{fin} = 9.1\times 10^7\ M_\odot$ and $\omega_\text{fin} = 3.92\ \text{Gyr}^{-1}$, we find that for $\hat{\lambda}_\text{ini} = 0$ the required unscaled dimensionless $\hat{\omega}_\text{ini} \approx 0.042$. The scaled lifetime of $\tau \approx 337.44\ t_\text{uni}$. For $\hat{\lambda}_\text{ini} = -0.05$, the corresponding self-coupling turns out to be $\lambda \approx -2.78\times 10^{-92}$ with a lifetime of $\tau \approx 8.74\times 10^4\ t_\text{uni}$. On the other hand, for $\hat{\lambda}_\text{ini} = 0.1$, the lifetime turns out to be $\tau \approx 5.9\times 10^{-2}\ t_\text{uni}$ with a scaled self-coupling $\lambda \approx 1.11\times 10^{-91}$. 
\end{itemize}

The impact of self-interactions we observe above agrees with the full numerical approach qualitatively, albeit the actual values of lifetimes are far from the ones we get from a numerical implementation. This is due to the approximate nature of this method, as discussed in the previous section. However, the intuition regarding the effect of scaling relations on the potential barrier experienced by the soliton remains intact, as seen in figure~\ref{fig:approx_SI_barriers} for both the fixed central density and fixed core mass cases.

\appendix

\end{document}